\documentclass[reprint,pra]{revtex4-2}
\usepackage{graphicx,amsmath,amsfonts}
\usepackage{dcolumn}
\usepackage{mathrsfs}
\usepackage{xcolor}
\usepackage{calligra}
\usepackage{bm}
\usepackage{enumerate}
\usepackage{pgfplots}
\usepackage{tikz}
\usetikzlibrary{decorations.markings}
\usetikzlibrary{plotmarks}
\usepackage[export]{adjustbox}

\newcommand{\n}{\hat{\mathbf{n}}}
\newcommand{\eps}{\boldsymbol{\varepsilon}}

\newcommand{\tp}{\left( t \right) }
\newcommand{\fb}{\mathbf{f} }

\newcommand{\mt}{\Tilde{m}}
\newcommand{\nt}{\Tilde{n}}
\newcommand{\st}{\Tilde{s}}

\newcommand{\Lgr}{\mathcal{L}}

\newcommand{\Ul}{\mathbf{U}^\parallel}
\newcommand{\Ua}{\mathbf{U}^a}
\newcommand{\Ub}{\mathbf{U}^b}
\newcommand{\Up}{\mathbf{U}^\parallel}
\newcommand{\Ut}{\mathbf{U}^\perp}
\newcommand{\Wl}{\mathbf{U}^\parallel}

\newcommand{\Ab}{\mathbf{A}}

\newcommand{\X}{\mathbf{X}}

\newcommand{\dAb}{\dot{\boldsymbol{A}}}
\newcommand{\dYb}{\dot{\boldsymbol{Y}}_{\nu}}
\newcommand{\ddYb}{\ddot{\boldsymbol{Y}}_{\nu}}
\newcommand{\Yb}{{\boldsymbol{Y}}_{\nu}}
\newcommand{\xib}{\mathbf{X}}
\newcommand{\dxib}{\dot{\mathbf{X}}}
\newcommand{\A}{\mathbf{A}}

\newcommand{\rb}{\mathbf{r}}
\newcommand{\rp}{\left( \mathbf{r} \right)}
\newcommand{\rpp}{\left( \mathbf{r}' \right)}
\newcommand{\dV}{\text{d}^3{\bf r}}
\newcommand{\dk}{\text{d}^3{\bf k}}
\newcommand{\dS}{\text{d}^2{\bf r}}

\makeatletter
\newsavebox{\@brx}
\newcommand{\llangle}[1][]{\savebox{\@brx}{\(\m@th{#1\langle}\)}%
  \mathopen{\copy\@brx\kern-0.5\wd\@brx\usebox{\@brx}}}
\newcommand{\rrangle}[1][]{\savebox{\@brx}{\(\m@th{#1\rangle}\)}%
  \mathclose{\copy\@brx\kern-0.5\wd\@brx\usebox{\@brx}}}

\colorlet{LightGray}{gray!10}
\makeatother

\definecolor{matlab1}{HTML}{0072BD}
\definecolor{matlab2}{HTML}{D95319}
\definecolor{matlab3}{rgb}{0.4940,0.1840,0.5560}

\newlength{\mywidth}
\newlength{\myheight}
\newlength{\mydelta}
\newif\iffig
\figfalse

\begin{document}


\title{Time-Domain Formulation of Electromagnetic Scattering \\Based on a Polarization Mode Expansion\\and the Principle of Least Action\\}
\author{Carlo Forestiere}
\affiliation{ Department of Electrical Engineering and Information Technology, Universit\`{a} degli Studi di Napoli Federico II, via Claudio 21,
 Napoli, 80125, Italy}
\author{Giovanni Miano}
\affiliation{ Department of Electrical Engineering and Information Technology, Universit\`{a} degli Studi di Napoli Federico II, via Claudio 21,Napoli, 80125, Italy}

\begin{abstract}
A fresh approach to the full wave analysis of time evolution of the polarization induced in the electromagnetic scattering from dispersive non magnetic particles is presented. It is grounded on the combination of the Hopfield model for the polarization field, the expansion of the polarization field in terms of static longitudinal and transverse modes of the particle, the expansion of the radiation field in terms of transverse wave modes of free space, and the principle of least action. The polarization field is linearly coupled to the electromagnetic field. The losses of the matter are provided through a linear coupling of the polarization field to a bath of harmonic oscillators with a continuous range of natural frequencies. The set of linear ordinary differential integral equations of convolution type of the overall system is reduced by eliminating both the radiation degrees of freedoms and the bath degrees of freedom, and the reduced system of equations is studied. The role played by the radiation field in the coupling between the longitudinal and transverse mode amplitudes of the polarization is described. The principal characteristics of the temporal evolution of the mode amplitudes are found as the particle size varies, including the impulse response. Results are presented for the analytically solvable spherical particle. The proposed approach leads to a general method for the analysis of the temporal evolution of the polarization field induced in dispersive particles of any shape, as well as for the computation of transients and steady states.
\end{abstract}

\maketitle


\section{Introduction}

The interaction of light with collective oscillations of matter in conducting and dielectric materials is one of the most active branches of optics: it enables the subwavelength confinement of electromagnetic fields and the enhancement of light-matter interaction (e.g., \cite{novotny_principles_2006},
\cite{koenderink_nanophotonics_2015}). The analysis and design of ultrafast devices requires the description of this interaction in the temporal domain (e.g., \cite{stockman_ultrafast_2008,shcherbakov_ultrafast_2015,mazzanti_all-optical_nodate}). The temporal evolution of linear time-invariant systems can be studied by using finite difference time domain methods or semi-analytical methods based on concepts such as natural mode, natural frequency, decay rate, impulse response and coupled modes.

Time-domain electromagnetic scattering by metal particles with arbitrary shapes has been investigated in terms of natural modes  using the quasi-electrostatic approximation and disregarding the coupling to radiation \cite{mayergoyz_analysis_2007}; radiative corrections have been evaluated by applying perturbative techniques \cite{Isaak_Rad_Corr}.  Time-domain electromagnetic scattering by spherical particles has been studied by using either the combination of the Mie theory, Debye series, and Fourier transform \cite{lock_mie_2011-1}, \cite{lock_mie_2011}, or solving time-domain integral equations \cite{li_time-dependent_2015}. The electromagnetic scattering in time domain from particles of arbitrary shape is studied by combining the Fourier transform with an expansion in terms of quasi-normal modes, \cite{lalanne_light_2018,faggiani_modal_2017,yan_rigorous_2018,Mod_open}, which are also called resonant states \cite{muljarov_brillouin-wigner_2010}.
The concept of quasi-normal mode traces back to the theory of natural electromagnetic oscillations, given by Stratton \cite{Stratton}. Time-domain modal expansions in the low frequency regime \cite{ammari_modal_2020,baldassari_modal_2021} have obtained by combining the Fourier transform and asymptotic analysis. The decay rate and the frequency shift of the plasmon modes in arbitrarily shaped metal nanoparticles have been studied by using the Fano-Hopfield approach and the pole approximation \cite{forestiere_quantum_2020}. In this paper we propose a different full wave approach to the study of the linear electromagnetic scattering from dispersive particles in the time domain: it is based on i) the separation of the degrees of freedom of the electromagnetic field from the induced polarization field; ii) a discrete expansion of the polarization field in terms of static longitudinal and transverse modes of the particle; iii) a continuum expansion of the radiation field in terms of transverse vector wave modes. The separation of the degree of freedom of matter and radiation field allows a detailed description of the field-matter interaction. The set of the static longitudinal and transverse modes of the particle is a basis for the space of the square integrable solenoidal vector fields defined on the region occupied by the particle.

By following the seminal works of U. Fano \cite{fano_atomic_1956} and J. J. Hopfield \cite{hopfield_theory_1958}, we represent the induced polarization as a vector field of coherent harmonic oscillators that is confined within the particle and it is linearly coupled to the electromagnetic field. The polarization field is also coupled linearly to a bath of harmonic oscillators with a continuous range of natural frequencies to describe phenomenologically the absorption of the matter \cite{huttner_quantization_1992}. Both the polarization density field and the bath field are expanded in terms of the static longitudinal and transverse modes of the particle. The electromagnetic field is represented in terms of the longitudinal component, i.e., the Coulombian field, and the transverse component, i.e., the radiation field. 
The longitudinal electric field is expressed in terms of the longitudinal degrees of freedom of the polarization field, while the transverse electric field is expanded in terms of transverse vector wave modes of the electromagnetic field in free space.
The equation of motion of the mode amplitudes are obtained by applying the principle of least action to the entire system ``polarization field + bath field + radiation field" by using the Lagrangian in the Coulomb gauge (e.g., \cite{cohen-tannoudji_photons_1997}). The system of equations governing the time evolution of the mode amplitudes of the polarization field are obtained by eliminating the degrees of freedoms of the bath field and of the radiation field. The coupling between the polarization and the radiation field determines the self- and mutual- interactions of the longitudinal and transverse mode amplitudes of the polarization. In general, the interaction between the amplitude of the mode $\Ua_p(\rb)$ and the amplitude of the mode $\Ub_{p'}(\rb)$ is expressed through a convolution integral with a kernel proportional $\int_{V}\dV\int_{V}\dV'\, \Ua_p(\rb)\overleftrightarrow{g}^\perp(\rb-\rb';t)\Ub_{p'}(\rb')$ where $\overleftrightarrow{g}^\perp(\rb-\rb';t)$ is the transverse Green function for the vacuum in the time domain. This double integral describes the exchange of electromagnetic energy between the two modes, which is a non-conservative process due to the electromagnetic energy radiated toward the infinity.

The expansion of the polarization density field in terms of the static longitudinal and transverse modes of the particle has several advantages: (i) the static longitudinal and transverse modes of the particle are an orthonormal basis for the space of square integrable solenoidal functions defined on the particle domain; (ii) the static longitudinal modes diagonalize the Coulombian energy contribution to the Lagrangian of the system. This greatly simplifies the system of equations governing the time evolution of the mode amplitudes of the polarization, because the coupling between the polarization degrees of freedom is limited to that due to the radiation field; (iii) the static transverse modes of the particle diagonalize the contribution of the singular term of $\overleftrightarrow{g}^\perp$ to the interaction integral of two generic transverse modes; (iv) as consequence of (ii) and (iii) the static longitudinal and the transverse modes of the particle are the natural modes (normal modes) of the polarization field in the small size limit; (v) although the radiation field couples the longitudinal and transverse mode amplitudes of the polarization, the coupling actually involves only a limited number of degrees of freedom, which depends on the ratio between the particle size and a characteristic wavelength of the material.

The paper is organized as follows. In Sec. \ref{sec:LongTrans} we present the longitudinal and transverse modes of a particle that we use to represent the polarization field. In Sec. \ref{sec:LagHam}, we introduce the Lagrangian of the system ``polarization field + bath field + radiation field", the principle of least action and the Lagrangian in the Coulomb gauge. In Sec. \ref{sec:Representation} we introduce the mode expansions for the matter field, the bath field and the radiation field. In Sec. \ref{sec:Lagrange} we apply the principle of least action to the Lagrangian expressed in terms of the mode amplitudes and obtain the Lagrange equations for the degrees of freedom of the entire system. In Sec. \ref{sec:EqMotion} we derive the system of ordinary differential - integral equations of convolution type governing the mode amplitudes of the polarization field. In Sec. \ref{sec:Results} we present the results for the analytically solvable spherical particle and we validate them. We discuss the main results and conclude in Sec. \ref{sec:Conclusion}.

\section{Static Longitudinal and Transverse Modes of the Particle}
\label{sec:LongTrans}
Throughout this manuscript, we denote with $V$ the region occupied by the particle, with $\partial V$ the boundary of $V$, with $\n$ the normal to the surface $\partial V$ pointing outward, and with $V_\infty$ the entire space. We introduce the scalar product 
\begin{equation}
    \langle {\bf C}, {\bf D} \rangle_W = \int_W {\bf C}^* \rp \cdot \mathbf{D} \rp \dV,
\end{equation}
and the norm $\left\| {\bf C} \right\|_W = \sqrt{ \langle {\bf C}, {\bf C} \rangle_W }$.  If the integration domain is not explicitly indicated, the scalar product is defined over $V$. 

In this paper, we consider the linear electromagnetic scattering by a dispersive, nonmagnetic, isotropic, and homogeneous particle. Due to the hypothesis of homogeneity, the induced polarization field is solenoidal in $V$, and its normal component on $\partial V$ is, in general, different from zero. 

The particle has two sets of orthogonal static  modes according to the scalar product $\langle {\bf C}, {\bf D} \rangle$ through which we can represent any square integrable solenoidal vector field defined on $V$, with non-zero normal component to $\partial V$. They are the \textit{static longitudinal} modes and the \textit{static transverse} modes of the particle. They only depend on the geometry of the particle, and are independent of the particle material. The static longitudinal modes are irrotational and solenoidal in $V$, but their normal component to $\partial V$ is different from zero; we denote them with the superscript $\parallel$. The static transverse modes are solenoidal in $V$, and their normal component to $\partial V$ is equal to zero; we denote them with the superscript $\perp$.

The static longitudinal (electrostatic) modes of the particle are solution of the eigenvalue problem \cite{fredkin_resonant_2003},\cite{forestiere_resonance_2020},
\begin{equation}
    \nabla \oint_{\partial V} \frac{ \Wl_p \rpp \cdot \n \rpp}{ 4 \pi \left| \rb - \rb' \right| } \dS' =\frac{1}{\lambda_p}\Wl_p \rp 
    \quad \text{in} \, V,
    \label{eq:EQS}
\end{equation}
where $\lambda_p$ is the eigenvalue associated to the eigenmode $\Wl_p\rp$. Apart from the factor $1/\varepsilon_0$, the integro-differential operator on the left-hand side of Eq. \ref{eq:EQS} gives the electrostatic field generated by a surface charge distribution as function of the surface charge density. The eigenvalues are discrete, real, positive, and equal or greater than two, $\lambda_p \ge 2$. The eigenmodes and the eigenvalues only depend on the shape of $V$, they do not depend on its size. Two eigenmodes $\Wl_{p'}$ and $\Wl_{p}$ associated to distinct eigenvalues are orthogonal according to the scalar product $\langle {\bf C}, {\bf D} \rangle$. The set $\left\{\Wl_p\rp\right\}$ is a base for the space of square integrable irrotational and solenoidal vector fields defined on $V$ that have a non-zero normal component on $\partial V$. The solution of the problem \ref{eq:EQS} can be obtained using the method outlined in Refs. \cite{mayergoyz_electrostatic_2005,mayergoyz_plasmon_2012}.

The static transverse (magnetostatic) modes of the particle are solution of the eigenvalue problem \cite{forestiere_electromagnetic_2019}, \cite{forestiere_magnetoquasistatic_2020}
\begin{equation}
\int_{V} \frac{ \mathbf{U}_p^\perp\rpp }{4 \pi \left| \mathbf{r}- \mathbf{r}' \right|} \dV' = \frac{1}{\kappa_p} \mathbf{U}_p^\perp\rp \quad \text{in}\, V,
\label{eq:MQS}
\end{equation}
with
\begin{equation}
 \mathbf{U}_p^\perp\rp \cdot \n \rp=0 \qquad \text{on} \, \partial V,
 \label{eq:boundaryMQS}
\end{equation}
where $\kappa_q$ is the eigenvalue associated to the eigenmode $\mathbf{U}_p^\perp$.
Apart from the factor $\mu_0$, the integral operator on the left-hand side of equation \ref{eq:MQS} gives the static magnetic vector potential in the Coulomb gauge generated by a volume current distribution as function of the current density field. Equation \ref{eq:MQS} with the constraint \ref{eq:boundaryMQS} holds in weak form in the functional space equipped with the inner product $\langle\mathbf{C}, \mathbf{D} \rangle$, and constituted by the vector fields that are solenoidal in $V$ and have zero normal component to $\partial V$. The eigenvalues are discrete, real and positive. Two eigenmodes $\mathbf{U}_{p'}^\perp(\mathbf{r})$ and $\mathbf{U}_p^\perp(\mathbf{r})$ associated to distinct eigenvalues are orthogonal according to the scalar product $\langle {\bf C}, {\bf D} \rangle$. The set $\left\{\mathbf{U}_p^\perp\rp\right\}$ is a base for the space of the square integrable solenoidal vector fields defined on $V$ with normal components to $\partial V$ equal to zero. Since equation \ref{eq:MQS} with the constraint \ref{eq:boundaryMQS} holds in weak form, the normal component on $\partial V$ of the static vector potential generated by $\mathbf{U}_p^\perp$ is in general different from zero.  The problem \ref{eq:MQS} can be solved by using standard tools of computational electromagnetism as outlied in Ref. \cite{forestiere_magnetoquasistatic_2020}.

The set of static longitudinal modes $\left\{\Wl_p\rp\right\}$ is orthogonal to the set of static transverse modes $\left\{\mathbf{U}_q^\perp(\mathbf{r})\right\}$ according to the scalar product $\langle {\bf C}, {\bf D} \rangle$. Any square integrable solenoidal vector field defined on $V$ can be represented by using both these sets of modes. 

In this paper, we expand the polarization density field $\mathbf{P}\left(t;\mathbf{r} \right)$ in the region $V$ as
\begin{equation}
\label{eq:xpar}
\mathbf{P}\left(t;\mathbf{r} \right) = \displaystyle\sum_p \left[p_p^\parallel(t) \Ul_p \rp + p_p^\perp(t) \mathbf{U}_p^\perp(\mathbf{r})\right],
\end{equation}
where $\left\{p_p^\parallel(t)\right\}$ are the longitudinal degrees of freedom of the polarization field, and $\left\{p_p^\perp (t)\right\}$ are the transverse degrees of freedom. The modes $\{\Ul_p\}$ and $\{\mathbf{U}_p^\perp\}$ are dimensionless quantities, normalized in such a way $\left\|{\Ul_p} \right\|=\left\|{\mathbf{U}_p^\perp} \right\|=1$.
We apply the principle of least action (e.g., \cite{cohen-tannoudji_photons_1997}) to obtain the equations governing the temporal evolution of the degrees of freedom of the polarization field, $\left\{p_p^\parallel(t)\right\}$ and $\left\{p_p^\perp (t)\right\}$.
\section{Formulation of the Electromagnetic Scattering Problem}
\label{sec:LagHam}

In this paper we describe the polarization field induced in the particle through the Hopfield model (\cite{fano_atomic_1956}, \cite{hopfield_theory_1958}). The polarization is represented as a continuum of harmonic oscillators with natural frequency $\omega_0$, linearly coupled to the electromagnetic field. The displacement vector field $\mathbf{X}\left(t; \mathbf{r} \right)$, denoted as the {\it matter field}, describes the continuum of harmonic oscillators; it is defined on $V$. The induced polarization density field $\mathbf{P}$ is related to the matter field $\mathbf{X}$,
\begin{equation}
\label{eq:Curr}
    \mathbf{P} \left(t;\mathbf{r} \right) = \left\{
    \begin{array}{cl}
        -\alpha_0 \xib\left(t;\mathbf{r} \right) & \text{in} \; V,  \\
          0 & \text{in} \; V_\infty \backslash V;
    \end{array}
    \right.
\end{equation}
where the parameter $\alpha_0$ determines the strength of the coupling between the electromagnetic field and the matter field; $\alpha_0$ has the dimension of an electric charge per unit of volume. The effects of the material losses are introduced through the phenomenological approach proposed by Huttner and Barnett \cite{huttner_quantization_1992} in which the matter field is also linearly coupled to a bath of harmonic oscillators with a continuous range of natural frequencies. The bath is described by the vector field $\Yb\left(t;\mathbf{r} \right)$, denoted as the \textit{bath\, field}, defined in $V$ and labeled by the frequency $\nu$. The coupling of the bath field $\Yb$ to the matter field $\mathbf{X}$ is described by a parameter $\upsilon(\nu)$. The bath field has the dimension of a length divided the square root of a frequency; $\upsilon_0$ has the dimension of mass per unit of volume times the square root of a frequency. The natural frequency $\omega_0$, the mass density of harmonic oscillators $\rho_0$, the coupling parameters $\alpha_0$ and $\upsilon(\nu)$ are assumed to be uniform in $V$. By choosing $\upsilon_0(\nu)/\rho_0 = \sqrt{2\gamma/\pi}$, $\alpha_0=en_0$ and $\rho_0=m_en_0$ ($e$ is the absolute value of the electron charge and $m_e$ is the electron mass) we obtain the Drude-Lorentz model with plasma frequency $\omega_P$, resonant frequency $\omega_0$ and damping rate $\gamma$. For dielectrics, $n_0$ is the number density of bound electrons and $\omega_0$ is the natural frequency of the bound electrons. For metals $n_0$ is the number density of free electrons and $\omega_0$ is equal to zero. The extension of the Hopfield model to include a generalized Lorentz description with multiple effective oscillators is possible.

\subsection{The Standard Lagrangian}

 The particle is excited by an incident electromagnetic field $\mathbf{E}_{inc}\left(t; \mathbf{r} \right)$ that is assumed to be equal to zero for $t<0$, thus the entire system is at rest at $t=0$. The field $\mathbf{E}_{inc}$ is solenoidal in $V$.

The field $\mathbf{X}\left(t; \mathbf{r}\right)$ is the natural generalized coordinate of the matter, and the field $\Yb\left(t; \mathbf{r}\right)$ is the natural generalized coordinate of the bath at frequency $\nu$. For the electromagnetic field the magnetic vector potential is a convenient generalized coordinate. We introduce the induced vector potential $\mathbf{A}\left(t; \mathbf{r}\right)$
 \begin{equation}
 \label{eq:potvet}
    \mathbf{A}\left(t; \mathbf{r}\right)= - \int_{0}^{t}\mathbf{E}\left(\tau; \mathbf{r}\right) d\tau,
\end{equation}
where $\mathbf{E}\left(t; \mathbf{r}\right)$ is the induced electric field; the incident vector potential is given by $\mathbf{A}_{inc}\left(t; \mathbf{r}\right)= - \int_{0}^{t}\mathbf{E}_{inc}\left(\tau; \mathbf{r}\right) d\tau$.

The degrees of freedom of the entire system ``matter + bath + electromagnetic field" are $\X(t;\rb)$, $\Yb(t;\rb)$ and $\mathbf{A}\left(t;\mathbf{r}\right)$. The variables $\rb$ and $\nu$ play the role a continuous indices for the degrees of freedom. The standard Lagrangian of the entire system $\Lgr$ is the sum of five terms: the matter term $\Lgr_m$, the electromagnetic field term $\Lgr_{em}$, the interaction term between the matter and the electromagnetic field $\Lgr_{em}^{int}$, the bath term $\Lgr_{bath}$, and the interaction term between the matter and the bath $\Lgr_{bath}^{int}$. We have \cite{huttner_quantization_1992}, \cite{Suttorp},
\begin{equation}
    \Lgr =  \Lgr_m + (\Lgr_{em} +\Lgr_{em}^{int}) + (\Lgr_{bath} +\Lgr_{bath}^{int}),
\end{equation}
where
\begin{subequations}
\begin{eqnarray}
\label{eq:Lmat}
     &&\Lgr_m (\xib,\dxib) = \int_V \dV\,\frac{\rho_0}{2}\left(\dxib ^2-\omega_0^2\xib^2\right),\\
     &&\Lgr_{em} (\Ab, \dAb) = \int_{V_\infty}\dV\left[ \frac{\varepsilon_0}{2} \dAb^2 - \frac{1}{2\mu_0}\left( \nabla\times \mathbf{A}\right)^2 \right],\quad\\
    &&\Lgr_{em}^{int}(\dxib, \Ab) = -\int_{V}\dV\,\alpha_0 \dxib \cdot (\Ab+\Ab_{inc}),\\
    \label{eq:Res}
     &&\Lgr_{bath}(\Yb,\dYb)=\int_V\dV\int_{0}^{\infty}d\nu\frac{\rho_0}{2}\left(\dYb^2 -\nu^2\Yb^2\right),\\
    \label{eq:Resint}
    &&\Lgr_{bath}^{int}(\xib,\dYb)=-\int_V\dV\int_{0}^{\infty}d\nu\,\upsilon(\nu)\xib\cdot\dYb;
\end{eqnarray}
\end{subequations}
 We are indicating the partial derivative with respect to time with a dot.
\subsection{The Principle of Least Action}
The Maxwell - Lorentz equations arise naturally from the principle of least action (e.g., \cite{cohen-tannoudji_photons_1997}). The action of the system in the time interval $(t_1,t_2)$ is
\begin{equation}
 \label{eq:potvet}
   \mathcal{S} = \int_{t_1}^{t_2}\Lgr \left( t \right) dt.
\end{equation}
One considers the variation of $\mathcal{S}$ when the matter field $\xib(t;\rb)$ is varied by a quantity $\delta\xib(t;\rb)$, the bath field is varied by a quantity $\delta\Yb(t;\rb)$, and the induced vector potential is varied by a quantity $\delta\A(t;\rb)$, where $\delta\xib(t;\rb)$, $\delta\Yb(t;\rb)$ and $\delta\A(t;\rb)$ are equal to zero at times $t_1$ and $t_2$. By requiring that $\mathcal{S}$ is extremal, $\delta S=0$, and by imposing that the induced vector potential is equal to zero at infinity, one gets the Lagrange equations governing the generalized coordinates of the system (e.g., \cite{Suttorp}),
\begin{equation}
    \label{eq:disp} 
     \ddot{\xib} + \omega_0^2\xib= \frac{\alpha_0}{\rho_0}(\dot{\A}+\dot{\A}_{inc})-\int_{0}^{\infty}\frac{\upsilon(\nu)}{\rho_0}\dYb d\nu  \quad \text{\;in } V,
\end{equation}
\begin{equation}
    \label{eq:res} 
    {\ddYb} +\nu^2\Yb= \frac{\upsilon(\nu)}{\rho_0}\dot{\xib}  \quad \text{in } V  \text{ and for } 0\le\nu<\infty,
\end{equation}
\begin{equation}
    \label{eq:Vpot}
    \ddot{\A} + c_0^2\nabla \times \nabla \times \A = \frac{1}{\epsilon_0}\left\{
    \begin{array}{cl}
        -\alpha_0 \dot\xib\left(t;\mathbf{r} \right) & \text{in} \; V  \\
          0 & \text{in} \; V_\infty \backslash V
    \end{array} \right.
\end{equation}
where $c_0$ is the light velocity in vacuum. The first equation governs the time evolution of the matter field, the second equation governs the time evolution of the bath field, and the third equation governs the induced vector potential. Since the external vector potential is solenoidal in $V$ from equations \ref{eq:disp}-\ref{eq:Vpot} it follows that the matter field, the bath field and the induced vector potential field are solenoidal in $V$, too. Furthermore, equation \ref{eq:Vpot} indicates that there is a surface polarization charge on $\partial V$ with surface density $\sigma \left(t;\mathbf{r} \right)$ given by
\begin{equation}
\label{eq:scharge}
    \sigma= -\alpha_0 X_n \quad \text{on} \, \partial V,
\end{equation}
where $X_n = \xib \cdot \n$.

Due to the intrinsic spatial inhomogeneity of the problem, the direct solution of the system of equations \ref{eq:disp}-\ref{eq:Vpot} in the time domain is very challenging. We overcome the problem in this way. We expand first the matter field, the bath field and the vector potential in terms of suitable sets of vector fields depending only on space (modal expansion), and then, we require that $\mathcal{S}$ is extremal in order to determine the equations governing the expansion coefficients, which only depend on time. In particular, we obtain the equations governing the time evolution of the expansion coefficient for the matter field in a closed form. They are ordinary differential - integral equations of convolution type, which can be studied and solved by using standard techniques. Once the matter field has been evaluated, the electromagnetic field can be evaluated inside the particle by using Equations \ref{eq:disp}, \ref{eq:res}, and outside the particle by using the electromagnetic potentials.

\subsection{Coulomb gauge Lagrangian}
The vector potential $\Ab(t;\mathbf{r})$ is defined on $V_\infty$. It is convenient to represent it as
\begin{equation}
\label{eq:coulgauge}
   \Ab = \Ab^\parallel + \Ab^\perp \text{\;in } V_\infty,
\end{equation}
 where $\Ab^\parallel(t;\mathbf{r})$ is the irrotational component of $\Ab$ and $\Ab^\perp(t;\mathbf{r})$ is the solenoidal component; $\Ab^\perp$ is the vector potential in the Coulomb gauge. The normal components of $\Ab^\perp$ is continuous across $\partial V$, while the normal component of $\Ab^\parallel$ is discontinuous. Indeed, from \ref{eq:Vpot} (and according to \ref{eq:scharge}) we obtain
\begin{equation}
\label{eq:gauss}
    \n\cdot(\dot{\A}^\parallel_{out}-\dot{\A}^\parallel_{in})=-\frac{1}{\eps_0}  \sigma \quad \text{on} \, \partial V,
\end{equation}
where ${\A}^\parallel_{out}$ denotes the value of $\Ab^\parallel$ on the external face of $\partial V$ and ${\A}^\parallel_{in }$ the value on the internal face of $\partial V$. The vector fields $\Ab^\parallel$ and $\Ab^\perp$ are orthogonal according to the scalar product $\langle {\bf C}, {\bf D} \rangle_{V_\infty}$.

Equation \ref{eq:gauss} allows to eliminate from the Lagrangian the longitudinal component of the vector potential, which is not a true degree of freedom of the system. By using the decomposition \ref{eq:coulgauge} and equation \ref{eq:gauss} we obtain the following expression for $(\Lgr_{em}+\Lgr_{em}^{int})$,

\begin{equation}
 \label{eq:lagr2}
    \Lgr_{em} + \Lgr_{em}^{int}  =  \Lgr_{c}+\Lgr_{rad}+ \Lgr_{rad}^{int},
\end{equation}
where
\begin{subequations}
\begin{eqnarray}
     \Lgr_{c} &=& - \int_{\partial V}\dS \int_{\partial V}\dS'\,\frac{\alpha_0^2}{2\varepsilon_0} \frac{X_n (t;\mathbf{r}) X_n (t;\mathbf{r'})}{4\pi|\mathbf{r}-\mathbf{r}'|}, \qquad \\
     \Lgr_{rad} &=& \int_{V_\infty} \dV\, \left[ \frac{\varepsilon_0}{2} (\dAb^\perp)^2 - \frac{1}{2\mu_0} \left(\nabla \times \mathbf{A^\perp} \right)^2 \right],\qquad \\
     \Lgr_{rad}^{int}&=-& \int_{V} \dV \,\alpha_0 \dxib \cdot (\Ab^\perp+\Ab_{inc}).
\end{eqnarray}
\end{subequations}
The term $-\Lgr_{c}$ is the Coulombian interaction energy between the surface polarization charges induced on $\partial V$, while $\Lgr_{rad}$ and $\Lgr_{rad}^{int}$ are the contributions due the radiation field.

\section{Representation of the vector fields $\xib$, $\Yb$ and $\Ab^\perp$ }
\label{sec:Representation}
We now introduce the expansion for the vector fields $\xib(t;\rb)$, $\Yb(t;\rb)$ and $\Ab^\perp(t;\rb)$ that we use in the paper to evaluate the expression of the Lagrangian of the entire system to which we apply the principle of least action.

\subsection{Matter field $\xib$ and bath field $\Yb$}
The vector fields $\xib$ and $\Yb$ are solenoidal in $V$, but their normal components to $\partial V$ are different from zero. We represent them by using the Helmholtz decomposition theorem for vector fields defined on a bounded region.

The matter field is represented as the sum of two terms $\xib$ = $\xib^\parallel$ + $\xib^\perp$ where: i) the longitudinal vector field $\xib^\parallel$ is irrotational and solenoidal in $V$ and its normal component to $\partial V$ is equal to $\xib \cdot \n$; ii) the transverse vector field $\xib^\perp$ is solenoidal in $V$ and its normal component on $\partial V$ equal to zero. The vector fields $\xib^\parallel$ and $\xib^\perp$ are orthogonal according to the scalar product $\langle {\bf C}, {\bf D} \rangle$. The vector field $\Yb$ is represented in the same way. 

The set of static longitudinal modes of the particle $\left\{\Wl_p\rp\right\}$, which are solutions of the eigenvalue problem \ref{eq:EQS}, is a basis for space of longitudinal vector fields defined on $V$. The set of static transverse modes of the particle $\left\{\mathbf{U}_p^\perp\rp\right\}$, which are the solutions of the eigenvalue problem \ref{eq:MQS}, is a basis for space of transverse vector fields defined on $V$. We choose them to represent the longitudinal and transverse components of $\xib$ and $\Yb$. As we will see, this choice turns out to be very appropriate because: (i) the static longitudinal  and  the transverse  modes  of the particle are the natural modes of polarization of the particle in the small size limit; (ii) a limited set of static longitudinal and transverse modes are needed for particles with size of the order of the characteristic wavelength of the material.

The fields $\xib^\parallel$ and $\Yb^\parallel$ are represented as
\begin{equation}
\label{eq:xpar}
\xib^\parallel(t;\rb) = \displaystyle\sum_p x_p^\parallel(t) \Ul_p \rp ,
\end{equation}
\begin{equation}
\label{eq:ypar}
\Yb^\parallel(t;\rb) =  \displaystyle\sum_n y_{\nu,p}^\parallel(t) \Ul_p \rp , 
\end{equation}
where $\left\{x_p^\parallel(t)\right\}$ are the longitudinal degrees of freedom of the matter field and $\left\{y_{\nu,p}^\parallel (t)\right\}$ are the longitudinal degrees of freedom of the bath field.
The fields $\xib^\perp$ and $\Yb^\perp$ are represented as
\begin{equation}
\label{eq:xperp}
\xib^\perp(t;\rb) = \displaystyle\sum_p x_p^\perp(t) \Ut_p \rp,
\end{equation}
\begin{equation}
\label{eq:yperp}
\Yb^\perp(t;\rb) = \displaystyle\sum_p y_{\nu,p}^\perp(t) \Ut_p \rp ,
\end{equation}
where $\left\{x_p^\perp(t)\right\}$ are the transverse degrees of freedom of the matter field and $\left\{y_{\nu,p}^\perp(t)\right\}$ are the transverse degrees of freedom of the bath field. We recall that the modes $\{\Ul_p\}$ and $\{\Ut_p\}$ are dimensionless quantities, normalized in such a way $\left\|{\Ul_p} \right\|=\left\|{\mathbf{U}_q^\perp} \right\|=1$. 

The polarization field degrees of freedom and the matter field degrees of freedom are related by 
\begin{equation}
\label{eq:dof_l}
p_p^\parallel=-\alpha_0x_p^\parallel,
\end{equation}
\begin{equation}
\label{eq:dof_t}
p_p^\perp=-\alpha_0x_p^\perp.
\end{equation}
In the following, we denote with $\mathbb{P}$ the set of values of the discrete index $p$.

\subsection{Radiation field $\Ab^\perp$}

We now introduce the basis for representing the vector field $\Ab^\perp(t;\mathbf{r})$, which is solenoidal everywhere in $V_\infty$. 

Let us consider the solutions of the eigenvalue problem
\begin{subequations}
\begin{eqnarray}
    \label{eq:vector_modes1}
     \nabla \times \mathbf{c}_h &=&k_h\mathbf{d}_h \qquad\text{in} \quad V_\infty, \\
     \label{eq:vector_modes2}
    \nabla \times \mathbf{d}_h &=&k_h\mathbf{c}_h \qquad\text{in} \quad V_\infty
\end{eqnarray}
\end{subequations}
that are regular at infinity. They are solutions of vector Helmholtz equation in free space. The eigenvalue $k_h$ is continuous, real and positive, $0< k_h <\infty$. The eigenfunctions are real and orthonormal according to the scalar product $\langle {\bf C}, {\bf D} \rangle_{V_\infty}$. They are a basis for the space of square integrable solenoidal vector fields defined in $V_\infty$. We now introduce the complex functions
\begin{equation}
\label{eq:aexp}
     {\bf f}_{(h,\pm)}(\mathbf{r})= \mathbf{c}_h(\mathbf{r}) \pm i \, \mathbf{d}_h(\mathbf{r}).
\end{equation}
We denote with the label $q$ the pair $(h,+)$ and with the label $-q$ the pair $(h,-)$. The functions ${\bf f}_q$ and ${\bf f}_{-q}$ are two independent solutions of the eigenvalue problem \ref{eq:vector_modes1}, \ref{eq:vector_modes2} with the same eigenvalue $k_h$, which we denote with $k_q$. The label $q$ runs over a continuous - discrete set, which we denote with $\mathbb{Q}$.

The set $\{\bf f\}_q$ is also an orthonormal basis for the space of square integrable solenoidal vector fields defined in $V_\infty$. We represent $\Ab^\perp(t;\rb)$ as:
\begin{equation}
\label{eq:aexp}
     \Ab^\perp(t;\rb) = \sum_q a_q(t) \mathbf{f}_q \rp,
\end{equation}
where $\left\{a_q(t)\right\}$ are the degrees of freedom of the radiation field. Since $ \Ab^\perp$ is real and $\mathbf{f}_q^* = \mathbf{f}_{-q}$, we have $a_q^*$ = $a_{-q}$. The vector fields $\{\mathbf{f}_q\}$ are dimensionless quantities. 

In this paper, we use both the transverse vector plane waves and the transverse vector spherical waves as a basis for representing $\Ab^\perp$, see Appendix A. In particular, we use the transverse plane waves to show that the interaction between the polarization degrees of freedom is governed by the transverse component of the Green function for the vacuum. We use the transverse spherical modes to study the responses of the spherical particle.

\section{Lagrange's Equations}
\label{sec:Lagrange}
We first evaluate the expression of the terms of the Lagrangian in the Coulomb gauge by using the expansions \ref{eq:xpar}-\ref{eq:yperp} and \ref{eq:aexp}. Then, we derive the Lagrange's equations for the degrees of freedom of the entire system.

\subsection{Expansion of the Lagrangian terms}
The expression of $\Lgr_m$ in terms of the degrees of freedom of the matter field is:
\begin{equation}
\label{eq:Lagrm}
    \Lgr_m = \sum_p \frac{\rho_0}{2}\left[\left(\dot{x}_p^{\parallel \, 2} - \omega_0^2 x_p^{\parallel \, 2} \right)+\left(\dot{x}_p^{\perp \, 2} - \omega_0^2 x_p^{\perp \, 2} \right)\right].
\end{equation}
The orthogonality of the polarization modes guarantees the diagonal structure of the matter term. 

Only the longitudinal component of the matter field contributes to the Coulombian term $\Lgr_c$. Its expression is:
\begin{equation}
\label{eq:Lagrc}
    \Lgr_c = -\sum_p \frac{\rho_0}{2}\Omega_p^2 x_p^{\parallel \, 2},
\end{equation}
where
\begin{equation}
    \Omega_p = \omega_P \sqrt{ \frac{1}{\lambda_p}},
\end{equation}
$\lambda_p$ is the eigenvalues associated to the $p-$static longitudinal mode of the particle, and 
\begin{equation}
     \omega_P=\sqrt{\frac{\alpha_0^2}{\rho_0\varepsilon_0}}
\end{equation}
is the ``plasma frequency" of the continuum of oscillators. The static longitudinal (electrostatic) modes of the particles diagonalize the Coulombian term of the Lagrangian. We recall that, in the Lorentz atomic model of a dielectric $\alpha_0 = -n_0e$, $\rho_0 = n_b m_e$, $\omega_P^2=n_be^2/\epsilon_0 m_e$ where $n_b$ is the number density of bound electrons contributing to the polarization. In a metal, the number density $n_b$ is replaced by the number density of free electrons. 

The expression of $\Lgr_{rad}$ in terms of the degrees of freedom of the radiation field is
\begin{equation}
\label{eq:Lagrr2}
    \Lgr_{rad} = \frac{\varepsilon_0}{2}\sum_q \left(\dot{a}_q^*\dot{a}_q  - \omega_q^2a_q^*a_q \right),
\end{equation}
where 
\begin{equation}
    \omega_q^2 = c_0^2 k_q^2.
\end{equation}
The orthogonality of the transverse vector waves used to represent the radiation field preserves the diagonal structure of the term $\Lgr_{rad}$.

The expression of the matter-radiation field coupling term $\Lgr_{rad}^{int}$ as function of the degrees of freedom of the matter field and of the radiation field is
\begin{eqnarray}
\;\nonumber
  \Lgr_{rad}^{int} &=&-\alpha_0 \displaystyle\sum_{p,q}\left( \langle\mathbf{U}_p^\parallel, \fb_q \rangle \,\dot{x}_p^\parallel + \langle\mathbf{U}_p^\perp, \fb_q \rangle \,\dot{x}_p^\perp \right)a_q\\
\qquad &-&\alpha_0 \displaystyle\sum_{p}\left( \langle\mathbf{U}_p^\parallel, \Ab_{inc}\rangle \, \dot{x}_p^\parallel + \langle\mathbf{U}_p^\perp, \Ab_{inc}\rangle \,\dot{x}_p^\perp\right).
\label{eq:Lagrint1}
\end{eqnarray}
We note that $\langle\Ua_p, \fb_q \rangle$ is the coefficient of the expansion of the vector field $\Ua_p$ in terms of transverse vector waves given in Eq. \ref{eq:planewave}.

The expression of the bath term $\Lgr_{bath}$ as function of the degrees of freedom of the bath field is
\begin{eqnarray}
\;\nonumber
  \Lgr_{bath} &=& \sum_p \frac{\rho_0}{2}\int_{0}^{\infty}\left(\dot{y}_{\nu,p}^{\parallel 2} - \nu^2{y}_{\nu,p}^{\parallel 2}\right)d\nu\\
\qquad &+&\sum_p \frac{\rho_0}{2}\int_{0}^{\infty}\left(\dot{y}_{\nu,p}^{\perp 2} - \nu^2{y}_{\nu,p}^{\perp 2}\right)d\nu,
\label{eq:Lbath}
\end{eqnarray}

Lastly, the expression of the interaction term $\Lgr_{int}^{bath}$ as function of the degrees of freedom of the bath field and of the matter field is:
\begin{equation}
\label{eq:Lagrint2}
    \Lgr_{bath}^{int} = -\sum_p\int_{0}^{\infty}\upsilon(\nu)\left(\dot{y}_{\nu,p}^\parallel{x}_p^\parallel+\dot{y}_{\nu,p}^\perp{x}_p^\perp\right) d\nu.
\end{equation}

\subsection{Lagrange's Equations}

We now apply the principle of least action to the Lagrangian in the Coulomb gauge
\begin{equation}
 \label{eq:lagr2}
    \Lgr =  \Lgr_m +\Lgr_{c}+\Lgr_{rad}+ \Lgr_{rad}^{int} + \Lgr_{bath} +\Lgr_{bath}^{int},
\end{equation}
where the terms on the right hand side are given by Expressions \ref{eq:Lagrm}, \ref{eq:Lagrc}, \ref{eq:Lagrr2}, \ref{eq:Lagrint1}, \ref{eq:Lbath} and \ref{eq:Lagrint2} as function of the degrees of freedom
$\left\{x_p^\parallel(t)\right\}$, $\left\{x_p^\perp(t)\right\}$, $\left\{y_{\nu,p}^\parallel (t)\right\}$, $\left\{y_{\nu,p}^\perp (t)\right\}$ and $\left\{a_q(t)\right\}$. By applying the principle of least action we obtain the set of Lagrange equations for the entire system, which are of the form
\begin{equation}
\label{eq:EqLagr1}
    \frac{d}{dt}\frac{\partial{\Lgr}}{\partial\dot u} -\frac{\partial{\Lgr}}{\partial{u}} = 0
\end{equation}
where $u = x_p^\parallel, \, x_p^\perp,\,  y_{\nu,p}^\parallel,\, y_{\nu,p}^\perp$ and $a_q$; the derivatives with respect to $  y_{\nu,p}^\parallel,\, y_{\nu,p}^\perp$ and $a_q$ are functional derivatives.

The equations governing the time evolution of $x_p^\parallel$ and $x_{p'}^\perp$, for any $p$ and $p'$ belonging to $\mathbb{P}$, are
\begin{equation}
    \label{eq:eqxparallel}
    \ddot{x}_p^\parallel +(\omega_0^2+\Omega_p^2) x_p^\parallel-\frac{\alpha_0}{\rho_0}\sum_{q}\langle \Up_p,\fb_{q}\rangle\dot{a}_{q}\\
    +\int_{0}^{\infty}\frac{\upsilon(\nu)}{\rho_0}\dot{y}_{\nu,p}^\parallel d\nu=f_p^\parallel
\end{equation}
\begin{equation}
    \label{eq:eqxperp}
    \ddot{x}_{p'}^\perp +\omega_0^2 x_{p'}^\perp-\frac{\alpha_0}{\rho_0}\sum_{q}\langle \Ut_{p'},\fb_{q}\rangle\dot{a}_{q}\\
    +\int_{0}^{\infty}\frac{\upsilon(\nu)}{\rho_0}\dot{y}_{\nu,{p'}}^\perp d\nu=f_{p'}^\perp
\end{equation}
where
\begin{equation}
\label{eq:Eqf}
      f_p^a(t)  =   \frac{\alpha_0}{\rho_0}\langle \Ua_p, \dot\Ab_{inc}\rangle,
\end{equation}
and $a=\parallel$, $\perp$. 

The equations governing the time evolution of the bath degrees of freedom $y_p^\parallel$ and $y_{p'}^\perp$, for any $p$ and $p'$ belonging to $\mathbb{P}$ and $0\le\nu<\infty$, are
\begin{equation}
\label{eq:Eqyparallel}
    \ddot{y}_{\nu,p}^\parallel +\nu^2{y}_{\nu,p}^\parallel-\frac{\upsilon(\nu)}{\rho_0}\dot{x}_p^\parallel=0,
\end{equation}
\begin{equation}
\label{eq:Eqyperp}
    \ddot{y}_{\nu,{p'}}^\perp +\nu^2{y}_{\nu,{p'}}^\perp-\frac{\upsilon(\nu)}{\rho_0}\dot{x}_{p'}^\perp=0.
\end{equation} 

The equations governing the degrees of freedom of the radiation field $a_q$, for any $ q$ belonging to $ \mathbb{Q}$, are
\begin{equation}
\label{eq:Eqa}
   \ddot{a}_q + \omega_q^2{a}_q+ \frac{\alpha_0}{\varepsilon_0}\displaystyle\sum_{p}\left(\langle\fb_q,\mathbf{U}_p^\parallel \rangle \,\dot{x}_p^\parallel + \langle\fb_q,\mathbf{U}_p^\perp \rangle \,\dot{x}_p^\perp\right) = 0.
\end{equation}

We highlight that the way the radiation field contributes to the dynamics of the matter field resembles in some sense that of the bath field. Indeed, being the entire system initially at rest, from equations \ref{eq:Eqyparallel}-\ref{eq:Eqa} we obtain:
\begin{equation}
\begin{aligned}
    \dot{y}_{\nu\,p}^{\parallel}(t) &= \frac{\upsilon(\nu)}{\rho_0}\int_{0}^{\infty} w_\nu \left( t - \tau \right) \dot{x}_p^{\parallel}(\tau) d\tau, \\
    \dot{y}_{\nu\,p}^{\perp}(t) &= \frac{\upsilon(\nu)}{\rho_0}\int_{0}^{\infty} w_\nu \left( t - \tau \right) \dot{x}_p^{\perp}(\tau) d\tau,
\end{aligned}
\label{eq:dyparallel}
\end{equation}

\begin{eqnarray}
\label{eq:da}
\nonumber
    \dot{a}_q(t)&=& -\frac{\alpha_0}{\varepsilon_0}\sum_{p}\int_{0}^{\infty} w_{\omega_q} \left( t - \tau \right) [\langle\fb_q,\mathbf{U}_p^\parallel\rangle \,\dot{x}_p^\parallel(\tau) 
    \\&+&\langle\fb_q,\mathbf{U}_p^\perp \rangle \,\dot{x}_p^\perp(\tau)] d\tau,
\end{eqnarray}
where
\begin{equation}
   w_\omega\tp = \theta \tp cos(\omega t),
\end{equation}
and $\theta \tp$ is the Heaviside function. 
\\
\section{Equations of Motion for the Degrees of Freedom of the Polarization Field}
\label{sec:EqMotion}
We now first reduce the set of Lagrange's equations obtained in the previous section to a system of equations governing the time evolution of the degrees of freedom of the matter field. Then, we apply the Laplace transform to reveal the essential features of the coupling between the longitudinal and transverse degrees of freedom of the polarization due to the radiation field.

\subsection{Time domain}

By substituting Equations \ref{eq:dyparallel} and \ref{eq:da} into the differential equations \ref{eq:eqxparallel} and \ref{eq:eqxperp}, we obtain the system of integro-differential equations of convolution type for the degrees of freedom of the matter field, for any $p$ and $p'$ belonging to $\mathbb{P}$),
\begin{widetext}
\begin{equation}
 \label{eq:xparallel2}
    \left[\ddot{x}_p^\parallel +(\gamma_{bath}*\dot{x}_p^{\parallel})+\omega_0^2 x_p^\parallel\right]+\Omega_p^2 x_p^\parallel+ \omega_P^2\sum_{p''} \left[(s_{pp''}^{\parallel \, \parallel}* \dot{x}_{p''}^\parallel) + (s_{pp''}^{\parallel\perp}*\dot{x}_{p''}^\perp)\right]=f_p^\parallel(t)\,,
\end{equation}
\begin{equation}
 \label{eq:xperp2}
    \left[\ddot{x}_{p'}^\perp +(\gamma_{bath}*\dot{x}_{p'}^{\perp})+\omega_0^2 x_{p'}^\perp\right]+ \omega_P^2\sum_{p''} \left[(s_{p'p''}^{\perp \, \parallel}* \dot{x}_{p''}^\parallel) + (s_{p'p''}^{\perp\perp}*\dot{x}_{p''}^\perp)\right]=f_{p'}^\perp(t),
\end{equation}
\end{widetext} 
where
\begin{equation}
 \label{eq:dBeta1}
    (g_1*g_2)(t)=\int_{0}^{\infty} g_1(t - \tau) g_2(\tau) d\tau
\end{equation}
is the convolution integral of $g_1(t)$ and $g_2(t)$,
\begin{equation}
 \label{eq:dBeta1}
    s_{pp'}^{a\,b}(t)=\sum_{q}\langle \Ua_p,\fb_{q}\rangle \langle \fb_{q}, \Ub_{p'}\rangle w_{\omega_q}(t)
\end{equation}
with $a,b=\parallel,\perp$, and
\begin{equation}
 \label{eq:dBeta1}
    \gamma_{bath}(t)=\int_{0}^{\infty}\left[\frac{\upsilon(\nu)}{\rho_0}\right]^2 w_{\nu}(t)d\nu.
\end{equation}
The degrees of freedom of the polarization field are proportional to the degrees of freedom of the matter field, thus Equations \ref{eq:xparallel2} and \ref{eq:xperp2} also describe the evolution of $p_p^\parallel(t)=-\alpha_0 x^\parallel _p$ and $p_p^\perp(t)=-\alpha_0 x^\perp _p$. The interaction of the polarization field with the radiation field couples Equations \ref{eq:xparallel2} and \ref{eq:xperp2}. 

In the first couple of square brackets (from the left) in equations \ref{eq:xparallel2} and \ref{eq:xperp2}, the first term describes the effects of inertia of the coherent oscillators representing the polarization field. The convolution integral describes the action of the bath field on the polarization, which phenomenologically accounts for the material losses. The simplest phenomenological model is obtained by choosing $\upsilon(\nu)/\rho_0 = \sqrt{2\gamma/\pi}$ where $\gamma$ is a constant representing the decay rate due to the material losses. In this case $\gamma_{bath}(t)=\gamma \delta(t)$, and in both the equations the expressions in the first square brackets return the Drude-Lorentz model. The third term describes the force keeping the electrons bound to the atom for dielectrics; it is equal to zero for metals. 

The fourth term (from the left) in equation \ref{eq:xparallel2} arises from the interaction of the polarization with the Coulombian electric field: it is not present in equation \ref{eq:xperp2}. Indeed, this interaction only involves the longitudinal degrees of freedom of the polarization field, and it does not couple them because the static longitudinal (electrostatic) modes diagonalize the Coulombian term of the Lagrangian $\Lgr_c$. The frequency $\Omega_p$ only depends on the material characteristics and the particle shape: it does not depend on the particle size. This interaction is responsible for the plasmonic oscillations in metals.

In the second couple of square brackets (from the left) in equations \ref{eq:xparallel2} and \ref{eq:xperp2} the convolution integrals, with kernels $\{s_{pp'}^{a\,b}(t)\}$, describe the interaction between the polarization degrees of freedom mediated by the radiation field. These interactions couple the longitudinal and transverse degrees of freedom of the polarization: each degree of freedom interacts with itself and with the other degrees of freedom. The interaction kernels $\{s_{pp'}^{a\,b}(t)\}$ depend on the particle size. Throughout the paper, we denote with $a$ the radius of the smallest sphere that encloses the particle, that is, its largest linear dimension. The amplitude of the interaction kernels scale as $(k_Pa)^2$ as the dimensionless parameter $k_Pa$ varies, where $k_P \equiv \omega_P/c_0$. As we will see, the parameter $ k_P a $ plays a very important role: the {\it plasma wavelength} $\lambda_P \equiv 2 \pi / k_P$ appears as the natural characteristic dimension to describe how the matter interacts with the electromagnetic field. 

We notice that the quantity $\dot{x}_{p}^\parallel(t)(s_{pp''}^{\parallel \, \parallel}* \dot{x}_{p''}^\parallel)(t)$ is proportional to the work per unit of time done on the longitudinal $p-$polarization mode by the transverse electric field generated by the longitudinal $p''$-polarization mode; the quantity $\dot{x}_{p}^\parallel(t)(s_{pp''}^{\parallel \, \perp}* \dot{x}_{p''}^\perp)(t)$ is proportional to the work per unit of time done on the longitudinal $p-$polarization mode by the transverse electric field generated by the transverse $p''-$polarization mode; and so on. The energy exchange between the polarization modes is a non-conservative process due to the electromagnetic energy radiated toward the infinity.
 
 The response of the polarization field can be either characterized by the natural modes of the system or by the impulse response of the mode amplitudes. The natural modes are the solutions of the system of equations \ref{eq:xparallel2} and \ref{eq:xperp2} with $f_p^\parallel(t)=f_{p'}^\perp(t)=0$. When $k_P a \ll 1$,  which we call {\it small size limit} throughout the paper, the coupling between the equations of system \ref{eq:xparallel2} and \ref{eq:xperp2} is weak, and the natural modes of the polarization are the static longitudinal and transverse modes of the particle (see Appendix D). The longitudinal natural modes arise from the interplay between the energy stored in the electric field and the energy stored in the polarization, while the transverse natural modes arise from the interplay between the energy stored in the magnetic field and the energy stored in the polarization. The self-interaction resulting from the coupling of each polarization discrete degree of freedom with the continuum degrees of freedom of the radiation field is responsible for the frequency shift and the radiative decay of the natural modes. In this case, the mutual interaction between the modes mainly transfers energy between them. For $k_Pa \sim 1$ the coupling between the longitudinal and transverse modes is important, and standing waves arise from the interplay between the energies stored in the electric field and in the magnetic field. The impulse response $h_{pp'}^{a b}(t)$, where $a,b=\parallel,\perp$ and $p$, $p'$ belong to $\mathbb{P}$, is the time evolution of the degree of freedom $p$ of the $a$-component of the polarization field when $f_{p'}^b(t) = \delta(t)$, the forcing terms of all the other degrees of freedom are set equal to zero, and the initial conditions of the degrees of freedom at $t=0^-$ are equal to zero. The impulse responses are linear combinations of the natural modes. They enables the direct determination of the forced evolution of the system. In this paper, we mainly study the impulse responses of the amplitudes of the polarization modes.
\subsection{Laplace domain}
 
The consequences of the coupling of the polarization with the radiation field can be better understood by studying equations \ref{eq:xparallel2} and \ref{eq:xperp2} in the Laplace domain. Let us indicate with $U \left( s \right)$ the Laplace transform of a function $u \left(t\right)$ (which is equal to zero for $t<0$), $U\left(s \right) = \int_{0}^{\infty} u\left(t\right)e^{-st} dt$. The region of convergence includes the imaginary axis because of the matter and radiation losses.

Since the entire system is initially at rest, equations \ref{eq:xparallel2} and \ref{eq:xperp2} give, respectively,
\begin{equation}
 \label{eq:xparallel5}
    \left(\frac{\omega_P^2}{\chi}+\Omega_p^2\right){X}_p^{\parallel}+\omega_P^2\sum_{p''} s\left(S_{pp''}^{\parallel \, \parallel} X_{p''}^\parallel + S_{pp''}^{\parallel\perp}X_{p''}^\perp\right)=F_p^\parallel,
\end{equation}
and
\begin{equation}
\label{eq:xperp5}
\frac{\omega_P^2}{\chi}{X}_{p'}^{\perp}+\omega_P^2\sum_{p''}s \left(S_{pp'}^{\perp\parallel} X_{p''}^\parallel + S_{{p'}p''}^{\perp\perp}X_{p''}^\perp\right)=F_{p'}^\perp,
\end{equation}
where: $X_{p}^\parallel(s)$ and $X_{p}^\perp(s)$ are the Laplace transform of $x_{p}^\parallel(t)$ and $x_{p}^\perp(t)$, respectively;
\begin{equation}
\label{eq:EqLa1}
    \Gamma_{bath}(s)=\int_{0}^{\infty}\left[\frac{\upsilon(\nu)}{\rho_0}\right]^2\frac{s}{s^2+\nu^2}d\nu
\end{equation}
is the Laplace transform of $\gamma_{bath}(t)$;
\begin{equation}
\label{eq:ExpS}
    S_{pp'}^{a\,b}(s)=\sum_{q} \langle \Ua_p, \fb_{q} \rangle \langle \fb_{q},\Ub_{p'}\rangle \frac{s}{s^2+c_0^2 k^2}
\end{equation}
is the Laplace transform of the interaction kernel $ s_{pp'}^{a\,b}$;
\begin{equation}
 \label{eq:chi}
    \chi(s) = \frac{\omega_P^2}{s^2 +s\Gamma_{bath}(s)+ \omega_0^2}
\end{equation}
is the susceptibility of the particle material. If $\upsilon(\nu)/\rho_0 = \sqrt{2\gamma/\pi}$, then $\Gamma_{bath}(s)=\gamma$, and we obtain the susceptibility of the Drude-Lorentz model for a dispersive material. The form of the system of equations \ref{eq:xparallel5} and \ref{eq:xperp5} is very important: we infer from them that the result we have obtained can be extended  to a particle with any susceptibility $\chi(s)$.

We denote with $\mathcal{H}_{pp'}^{a b}(s)$ the Laplace transform of the impulse response $h_{pp'}^{a b}(t)$. Since the region of convergence of the Laplace transform contains the imaginary axis, we can evaluate the impulse response by performing the inverse Fourier transform of the frequency response $H_{pp'}^{a b}(\omega)=\mathcal{H}_{pp'}^{a b}(s=i\omega+\epsilon)$ where $-\infty<\omega<+\infty$ and $\epsilon \downarrow 0$. In this way we can easily manage the improper integrals with respect to the wavenumber $k$ appearing in the expression of the coefficients  $S_{p\,p'}^{a\,b}(i\omega+\epsilon)$ by using the relation $1/(x-i\epsilon)=i\pi\delta(x)+P_f(1/x)$ where $P_f$ denotes the principal value (for more details see Appendix \ref{sec:Coefficients}). The Dirac function contribution gives the imaginary terms that describe the radiation losses.

\subsection{Physical meaning of the coupling coefficient $S_{pp'}^{a\,b}(s)$}
In the Laplace domain the interaction kernels $\{S_{pp'}^{a\,b}(s)\}$ are meromorphic functions of $s$ with infinite number of poles, which are connected to the delay due to the non-zero size of the particle. This clearly emerges by using the transverse plane waves as the basis for the radiation field. We rewrite expression \ref{eq:ExpS} as
\begin{equation}
\label{eq:Sint}
   S_{pp'}^{a\,b}(s)=\frac{s}{c_0^2} \int_{V}\dV\int_{V}\dV'\, \Ua_p(\rb)\overleftrightarrow{G}^\perp(\rb-\rb';s)\Ua_{p'}(\rb'),
\end{equation}
where $\overleftrightarrow{G}^\perp(\rb-\rb';s)$ is the dyad
\begin{equation}
\label{eq:ExpS1}
    \overleftrightarrow{G}^\perp(\rb-\rb';s)=\sum_{q}\frac{1}{k^2+s^2/c_0^2} \fb_{q}(\rb)\otimes\fb_{q}^*(\rb').
\end{equation}
By using the expression \ref{eq:planewave} for $\fb_{q}(\rb)$ we have
\begin{equation}
\label{eq:ExpS1}
    \overleftrightarrow{G}^\perp(\rb;s)=\frac{1}{(2\pi)^3}\int{}\overleftrightarrow{\mathcal{G}}^\perp(\mathbf{k};s)\,e^{i\mathbf{k}\cdot\rb} d^3\mathbf{k}
\end{equation}
where
\begin{equation}
\label{eq:FourierI}
    \overleftrightarrow{\mathcal{G}}^\perp(\mathbf{k};s)= \frac{1}{k^2+s^2/c_0^2} (\overleftrightarrow{I}-\hat{\mathbf{k}}\otimes\hat{\mathbf{k}})
\end{equation}
is the transverse Green function for the vacuum in the wavenumber domain. By evaluating the Fourier integral \ref{eq:ExpS1} we obtain \cite{arnoldus_transverse_2003}
\begin{eqnarray}
\label{eq:Green}
\nonumber
    \overleftrightarrow{G}^\perp \left(\rb;s \right)=(\overleftrightarrow{I}-\hat{\mathbf{r}}\otimes\hat{\mathbf{r}})\frac{1}{4\pi r}e^{-sr/c_0}+
   \\
    (\overleftrightarrow{I}-3\hat{\mathbf{r}}\otimes\hat{\mathbf{r}})\frac{c_0^2}{4\pi sr^2}\left[\frac{1}{c_0}e^{-sr/c_0}-\frac{1}{sr}(1-e^{-sr/c_0})\right]. \quad
\end{eqnarray}
This is the full wave transverse dyadic Green function for the vacuum in the Laplace domain. The expression of the corresponding function in the time domain is immediate, the factor $e^{-sr/c_0}$ gives a
the retardation $r/c_0$.
Equations \ref{eq:Sint} and \ref{eq:Green} are very useful to evaluate numerically $S_{pp'}^{a\,b}(s)$ in the frequency domain for arbitrarily shaped particles. 

The interaction integrals $S_{pp'}^{a\,b}(s)$ describe the energy exchange between the particle polarization modes that is mediated by the radiation field, which is represented by the full wave transverse dyadic Green function for the vacuum. Since the longitudinal and transverse modes are normalized, the amplitudes of $\omega_P^2S_{pp'}^{a\,b}(s)$ scales according  to the dimensionless parameter $(k_P a)^2$ as $k_Pa$ varies.

\subsection{Natural modes of polarization in the small size limit}

Although in this paper we mainly study the impulse responses of the degrees of freedom of the polarization, for various important reasons we also analyze the natural modes of the polarization in the small size limit. 
In the Laplace domain the natural modes of the polarization field are the solutions of the nonlinear eigenvalue problem
\begin{eqnarray}
 \label{eq:xparallelom1}
  \nonumber
    \left[\frac{\omega_P^2}{\chi(\zeta)} + \Omega_p^2\right]{Z}_p^{\parallel}&+&
    \\
    \omega_P^2\sum_{p''} \zeta\left[S_{pp''}^{\parallel \, \parallel}(\zeta) Z_{p''}^\parallel + S_{pp''}^{\parallel\perp}(\zeta)Z_{p''}^\perp\right]&=&0, \quad
\end{eqnarray}
and
\begin{eqnarray}
\label{eq:xperpom1}
\frac{\omega_P^2}{\chi(\zeta)}{Z}_{p'}^{\perp}+\omega_P^2\sum_{p''} \zeta\left[S_{{p'}p''}^{\perp\, \parallel}(\zeta) Z_{p''}^\parallel + S_{{p'}p''}^{\perp\perp}(\zeta)Z_{p''}^\perp\right]&=0 \qquad
\end{eqnarray}
where $\zeta$ is the eigenvalue and ($Z_{p_1}^\parallel, $, $Z_{p_2}^\parallel$, ...,  $Z_{p'_1}^\perp$, $Z_{p'_2}^\perp$,...) are the components of the corresponding eigenvector. It can be reduced to a linear eigenvalue problem by introducing auxiliary degrees of freedom for taking into account both the dynamics of the matter and the radiation effects. To do this, we need the poles and the corresponding residuals of the susceptibility $\chi(\zeta)$ and of the coefficients $S_{pp'}^{a\,b}(\zeta)$. There are techniques that allow the approximation of meromorphic functions through rational functions with a finite number of dominant poles. In particular, the vector fitting technique \cite{gustavsen_improving_2006,gustavsen_rational_1999,deschrijver_macromodeling_2008} allows to calculate the dominant poles and residues from the frequency responses. The resulting approximation guarantees stable poles that are real or come in complex conjugate pairs, and the model can be directly converted into a state-space model.

The coupling  between the degrees of freedom of the polarization field in equations \ref{eq:xparallelom1}-\ref{eq:xperpom1} (in equations \ref{eq:xparallel5}-\ref{eq:xperp5}, and in equations \ref{eq:xparallel2}-\ref{eq:xperp2}) is weighted by the dimensionless size parameter of the particle $\beta=k_P a$. In the limit $\beta\rightarrow 0$ the system of equations \ref{eq:xparallelom1}-\ref{eq:xperpom1} reduces to a system of uncoupled equations (see Appendix E). Thus, the static longitudinal (electrostatic) modes and the static transverse (magnetostatic) modes of the particle are the natural oscillation modes of the polarization for $\beta\ll1$.
The transverse natural modes are degenerate because they have the same natural frequency for $\beta=0$. Nevertheless, this degeneration disappears by taking into account the non-zero size of the particle. By retaining the leading order terms in $\beta$ we obtain, respectively, for the natural frequency $\Omega^\parallel_p$ of the longitudinal $p-$mode, and for the natural frequency $\Omega^\perp_p$ of the transverse $p-$mode (see Appendix E),
\begin{equation}
\Omega^\parallel_p\cong\sqrt{\omega_0^2 +\Omega_p^2}\,(1-\beta^2 R_{pp}^{\parallel \parallel}),
\label{eq:PlasmonSmallResonances1}
\end{equation}
and
\begin{equation}
\Omega^\perp_p\cong\omega_0\left(1-\frac{\beta^2}{2a^2\kappa_p}\right).
\label{eq:DielectricSmallResonance1}
\end{equation}
The parameter $1/(a^2\kappa_p)$ does not depend on the size of the particle, it only depends on the shape. 

The choice to expand the polarization field in terms of the static longitudinal and transverse modes of the particles turns out to be very appropriate to describe the electromagnetic scattering from dispersive particles with size of the order of $1/k_P $ because they are the natural modes of polarization in the small size limit $k_Pa \ll1$. The static longitudinal modes of the particles are natural modes of polarization in the small size limit because they diagonalize the Coulombian term of the Lagrangian. Likewise, the static transverse modes of the particles are natural modes of polarization in the small size limit because in this limit they diagonalize the interaction matrix with coefficients $S_{pp'}^{\perp\,\perp}$. 

\section{Results for spherical particles}
\label{sec:Results}
In this Section, we apply the proposed approach to the case of a dispersive spherical particle of radius $a$, which can be solved semi-analytically. In particular, we focus on the effect of coupling between polarization and radiation fields as the dimensionless size parameter $\beta=k_Pa$ varies (where $k_P=\omega_P/c_0$), disregarding the material losses.
\subsection{Static longitudinal and transverse modes
of a sphere}
\begin{figure}
    \centering
    \includegraphics[width=\columnwidth]{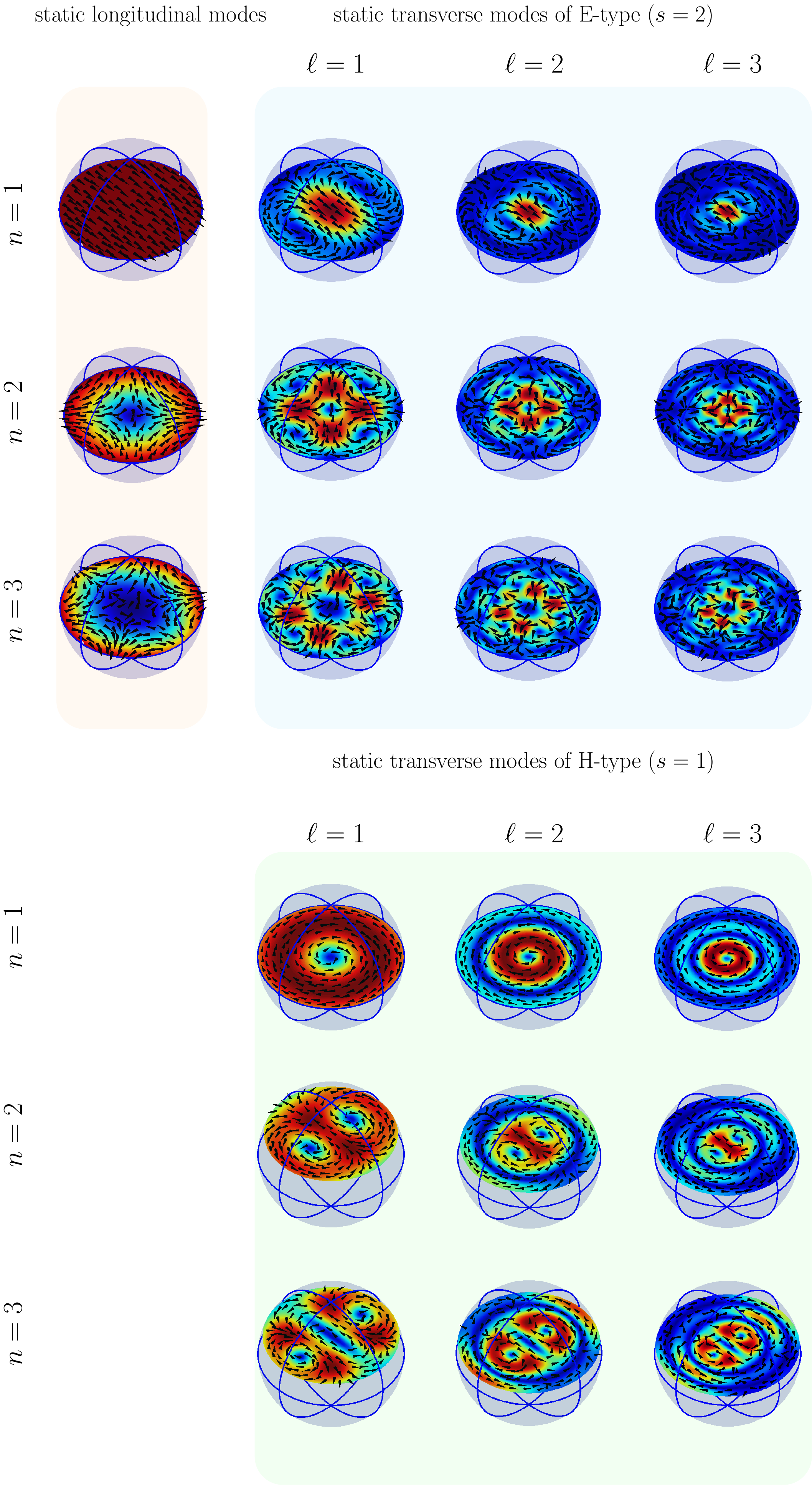}
    \caption{Static longitudinal modes of a sphere with multipolar order $n=1,2,3$, i.e., the electric dipole, quadrupole, and octupole. The static transverse modes of a sphere are divided into two subsets: the ones of {\it E-type} (with $s=2$), and the ones of {\it H-type} (with $s=1$), $n=1,2,3$ and $\ell=1,2,3$ \cite{bohren_absorption_1998}. Modes located on the some row interact due to the radiation coupling.}
    \label{fig:Modes}
\end{figure}
\begin{figure}
    \centering
    \includegraphics[width=\columnwidth]{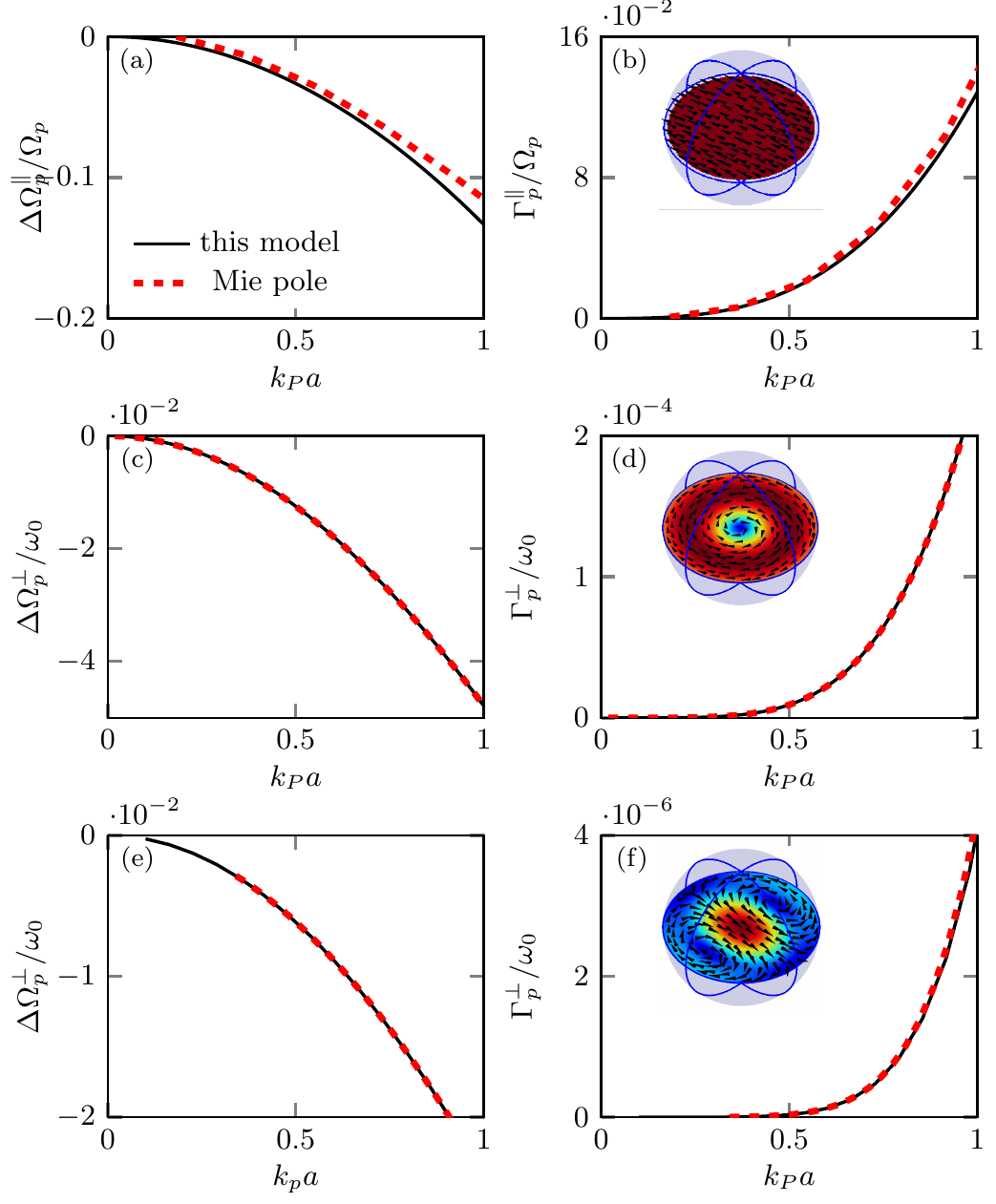}
    \caption{Spherical particle with radius $a$. Frequency shift (a),(c),(e) and radiative decay rate (b), (d), (f) of the longitudinal and transverse natural modes of the particle, normalized to their natural frequencies in the small size limit, as a function of $\beta =k_Pa$ where $k_P=\omega_P/c_0$; longitudinal mode with $p=\left(m1v\right)$ (electric dipole) in a metal sphere (a-b) ($\omega_0=0$); transverse H-type mode with $p=\left(mn1lv\right)$ (magnetic dipole) (c-d) and transverse E-type mode with $p=\left(mn2lv\right)$ (e-f) in a dielectric sphere with $\omega_0 = \omega_p/4$. Two different approaches have been used: the one proposed in this manuscript (full line) and the poles of the Mie coefficients (dashed line).}
\label{fig:comparison}
\end{figure}
The static longitudinal and transverse modes of a sphere have the analytical expressions given in Appendix \ref{sec:B}. In Fig. \ref{fig:Modes} we show the longitudinal modes with multipolar order $n=1,2,3$, namely the electric dipole, quadrupole, and octupole. The transverse modes are divided into two subsets: the ones of {\it E-type} and the ones of {\it H-type} \cite{bohren_absorption_1998}. In Fig. \ref{fig:Modes} we show the ones with $n=1,2,3$ and $\ell=1,2,3$. 

In the small size limit, these modes are the natural modes of the polarization of a spherical particle. The longitudinal modes resonate in a metal sphere at the frequency given by Equation \ref{eq:PlasmonSmallResonances1}. The transverse modes resonate in a dielectric sphere at the frequency given by Equation \ref{eq:DielectricSmallResonance1}, while they are off resonance in a metal sphere. Although these modes are orthogonal, they interact through radiative coupling due to the non-zero size of the particle. Their interaction properties are characterized by the coefficient $S_{pp'}^{a\,b}(s)$ evaluated in Appendix \ref{sec:Coefficients} for $s=i\omega+\epsilon$ in the limit $\epsilon \downarrow 0$. Here, we summarize the main properties: i) each mode interacts with itself; ii) only modes with the same multipolar order $n$ may interact; iii) transverse modes of H-type do not interact with either longitudinal modes or transverse modes of the  E-type ; iv) longitudinal modes and transverse modes of E-type interact.  This is exemplified in Fig. \ref{fig:Modes}, where the modes located on same row interact. 

\subsection{Frequency shift and decay rate of the longitudinal and transverse natural modes, and validation}

In the small size limit, the natural modes of polarization are the static longitudinal and transverse modes of the particle. Due to the non-zero size effects, the values of the natural frequencies of these modes deviate from those obtained in the limit $\beta\rightarrow{}0$. Furthermore, the modes amplitudes decay exponentially due to the radiation losses. We denote with $\Delta\Omega_p^a$ the difference between the non-zero size natural frequency of the $p$-mode of $a$ type and the value obtained for $\beta=0$ (frequency shift of the natural frequency) where $a=\parallel,\perp$; with $\Gamma_p^a$ we denote the radiative decay rate.

The frequency shift and the radiative decay rate of the longitudinal and transverse polarization modes can be evaluated approximately by using the pole approximation technique \cite{forestiere_quantum_2020}. It consists in approximating the coupling coefficients $S_{pp'}^{a\,b}(s)$ with their values at $s=i\Omega+\epsilon$ where $\Omega$ is the natural frequency of the mode for $\beta=0$ and $\epsilon \downarrow 0$. In this paper we evaluate the frequency shifts and the radiative decay rates by using the asymptotic expansions of the interaction coefficients given in Appendix \ref{sec:Asymptotic}, and by solving perturbatively the eigenvalue problem \ref{eq:xparallelom1} and \ref{eq:xperpom1}. We obtain  approximated analytical expressions for $0\leq\beta<1$. In Fig. \ref{fig:comparison}, we compare them against the corresponding quantities obtained from the poles of the Mie coefficients as a function of $\beta$. In panels (a-b) we consider the longitudinal mode $\mathbf{U}_{m1v}^\parallel$ with $n=1$ (electric dipole) of a metal particle (where $\omega_0=0$). In panels (c-d) we consider the H-type transverse mode $\mathbf{U}_{m111v}^\perp$ with $n=1$, $s=1$, and $\ell=1$ (magnetic dipole) for a dielectric particle with $\omega_0 = \omega_P/4$. Eventually, in panels (e-f) we consider the E-type transverse mode $\mathbf{U}_{m121v}^\perp$ with $n=1$, $s=2$ and $\ell=1$ (also known as {\it toroidal} dipole) for a dielectric particle with $\omega_0 = \omega_P/4$. Good agreement is found in any case. 

In the following, we give the asymptotic expressions of $\Delta\Omega_p^a$ and $\Gamma_p^a$ for $\beta\ll1$. For a metal spherical particle (where $\omega_0=0$) the frequency shift and the radiative decay rate of the longitudinal mode $p=\left(mnv\right)$ are given by
\begin{equation}
\frac{\Delta\Omega_p^\parallel}{\Omega_p}\cong-\frac{(n+1)}{(3+2 n)(4 n^2-1)}\beta^2,
\end{equation}
\begin{equation}
\frac{\Gamma_p^\parallel}{\Omega_p}\cong\frac{\left(n+1\right) (2n+1)}{n \left[\left(2 n+1\right) !!\right]^2}  \left(\frac{n}{2n+1}\right)^{(n + 1/2)} \beta^{ 2 n + 1 };
\end{equation}
here $\Omega_p$ denotes the natural frequency of the mode for $\beta=0$ and $\omega_0=0$. For dielectric particles, the frequency shift and the radiative decay rate of the transverse mode of H-type $p=\left(mn1lv\right)$ are given by
\begin{equation}
\frac{\Delta\Omega_p^\perp}{\omega_0}\cong-\frac{1}{2z_{n-1,\ell}^2}\beta^2,
\label{eq:DielectricSmallResonance2}
\end{equation}
\begin{equation}
\frac{\Gamma_p^\perp}{\omega_0} \cong \frac{2}{[ \left( 2 n-1 \right) ! !]^{2}}\frac{1}{ z_{n-1,\ell}^4}\left(\frac{ \omega_0}{\omega_P}\right)^{2n+1} \beta^{2n+3};
\end{equation}
where $z_{n-1,\ell}$ is the $\ell-th$ zero of the spherical Bessel function of order $n-1$. For the transverse mode of E-type with $p=\left(mn2lv\right)$ they are given by
\begin{equation}
\frac{\Delta\Omega_p^\perp}{\omega_0}\cong-\frac{1}{2z_{n,\ell}^2}\beta^2,
\label{eq:DielectricSmallResonance2}
\end{equation}
\begin{equation}
 \frac{\Gamma_p^\perp}{\omega_0} \cong \frac{2}{n^{2}[(2 n-1) ! !]^{2}}\frac{1}{ z_{n,\ell}^4} \left(\frac{ \omega_0}{\omega_P}\right)^{2n+3} \beta^{2n+5}
\end{equation}
where $z_{n,\ell}$ is the $\ell-th$ zero of the spherical Bessel function of order $n$. 

\begin{figure*}
    \centering
    \includegraphics[width=\textwidth]{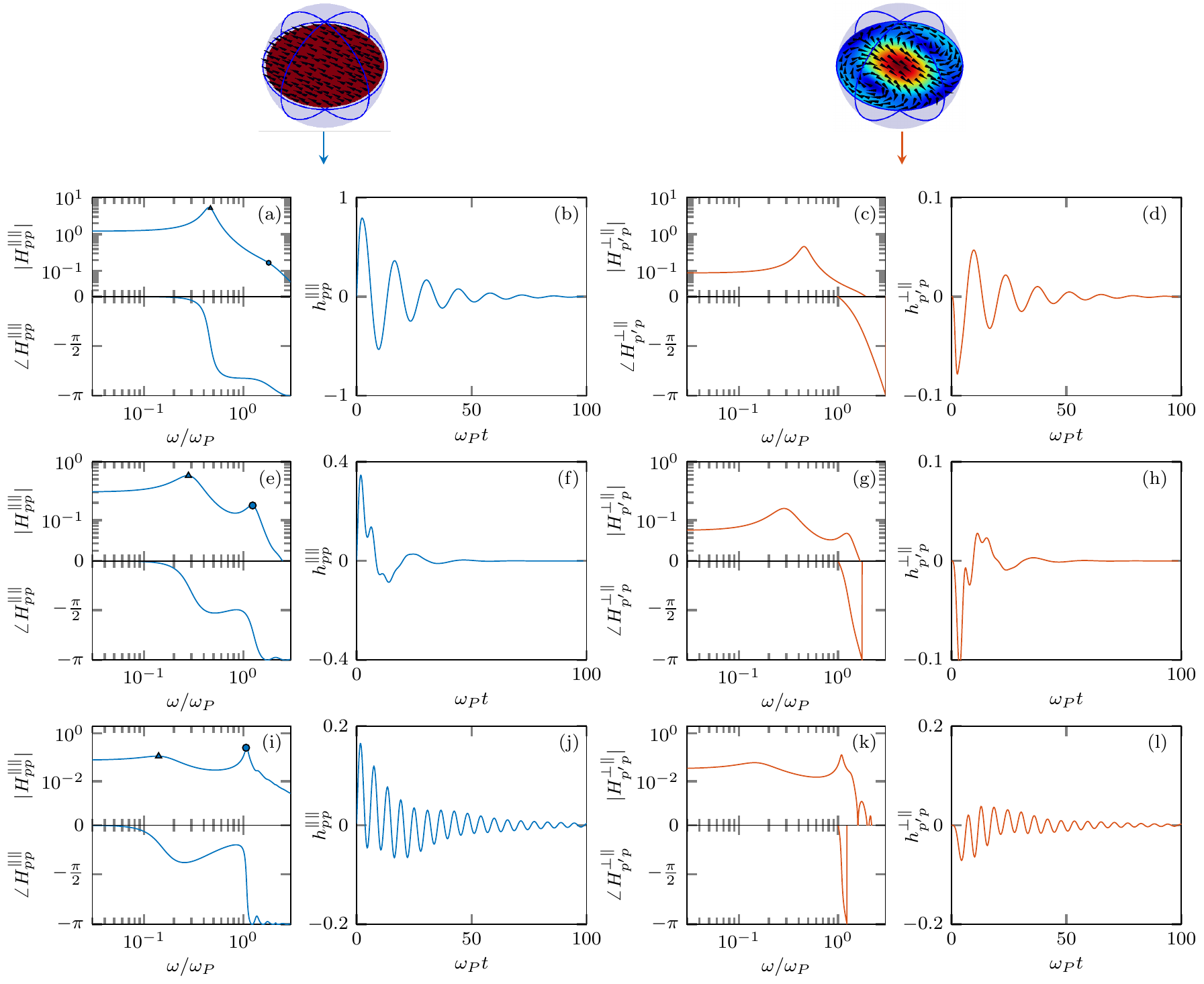}
    \caption{Frequency response $H^{\parallel \parallel}_{pp}$ (first column) and impulse response $h^{\parallel \parallel}_{pp}$ (second column) of the longitudinal mode amplitude with $p=\left(m1\nu\right)$ and forcing term $F_p^\parallel=1$; frequency response $H^{\perp \parallel}_{p'p}$ (third column) and impulse response $h^{\perp \parallel}_{p'p}$ (fourth column) of the transverse mode amplitude with $p' = \left(m121v\right)$. We consider a lossless metal sphere ($\omega_0 =0$, $\Gamma=0$) with $\beta=\pi/2$ (a-d), $\beta=\pi$ (e-f) and $\beta=2\pi$ (i-l), where $\beta=k_Pa$, $a$ is the radius and $k_P=\omega_P/c_0$.}
    \label{fig:LongLong}
\end{figure*}
\begin{figure}
    \centering
    \includegraphics[width=\columnwidth]{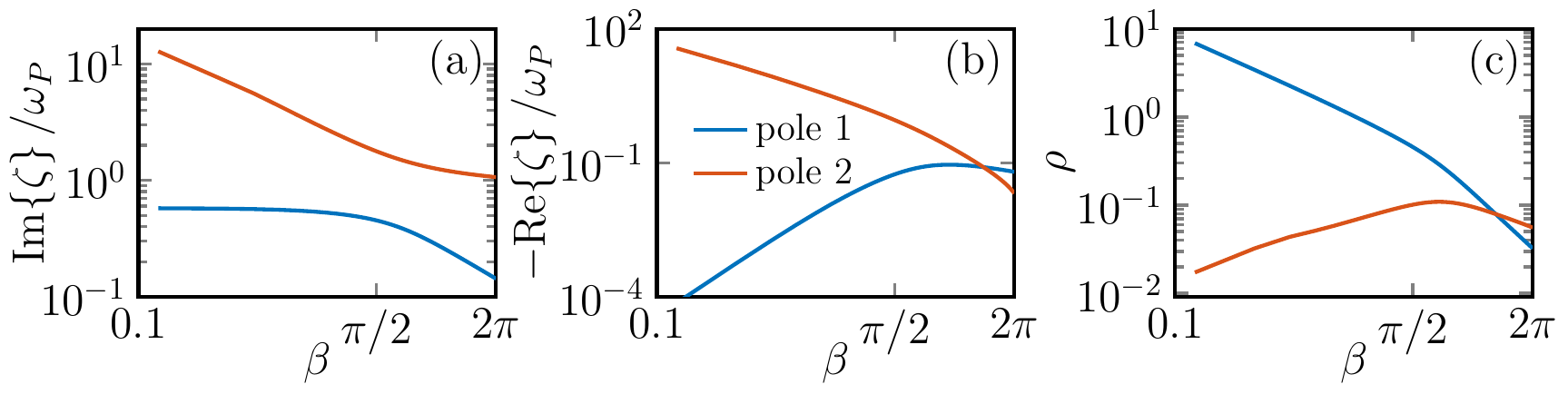}
    \caption{Imaginary (a) and minus real (b) parts of the first two dominant poles, with the corresponding magnitude of the residues (c), of the frequency response $H^{\parallel \parallel}_{pp}$ with $p=\left(m1\nu\right)$ of a lossless metal sphere ($\omega_0=0$, $\Gamma=0$) as a function of $\beta=k_Pa$ where $a$ is the radius and $k_P=\omega_P/c_0$. These quantities are obtained by using the vector fitting technique.}
    \label{fig:Vector_Fit}
\end{figure}
\begin{figure*}
    \centering
    \includegraphics[width=\textwidth]{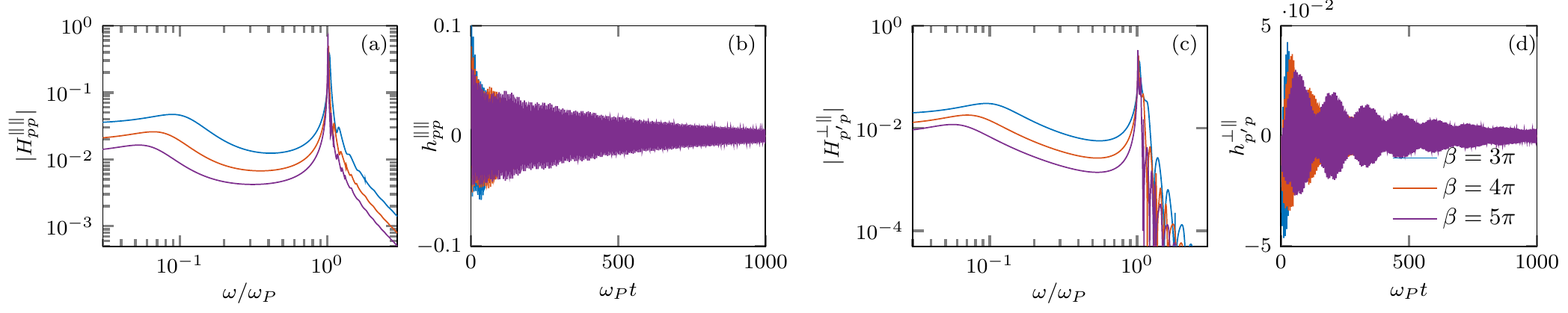}
    \caption{Amplitude response $|H^{\parallel \parallel}_{pp}|$ (a) and impulse response $h^{\parallel \parallel}_{pp}$ (b) with $p=\left(m1\nu\right)$ and forcing term $F_p^\parallel=1$; amplitude response $|H^{\perp \parallel}_{p'p}|$ (c) and impulse response $h^{\perp \parallel}_{p'p}$ (d) with $p' = \left(m121v\right)$. We consider a lossless metal sphere ($\omega_0 =0$, $\Gamma=0$) with  $\beta=3\pi$, $\beta=4\pi$ and $\beta=5\pi$, where $\beta=k_Pa$, $a$ is the radius and $k_P=\omega_P/c_0$.}
    \label{fig:RadiativeRegime}
\end{figure*}
\subsection{Frequency and impulse response of the mode amplitudes in a metal particle}

We now investigate the response of a metal particle ($\omega_0=0$) beyond the small size limit, as $\beta=k_Pa$ varies in the interval $\left[0,2\pi\right]$. The longitudinal modes and the transverse modes of E-type with the same multipolar order are coupled through the radiation field. In particular, we consider the response of the amplitude of the longitudinal mode $\mathbf{U}_{p}^\parallel$ with $p=\left(m1v\right)$, and the response of the amplitude of the transverse mode of E-type $\mathbf{U}_{p'}^\perp$ with $p'=\left(m121v\right)$. In the system of equations \ref{eq:xparallel5}, \ref{eq:xperp5} (\ref{eq:xparallel2}, \ref{eq:xperp2}) we have set $F_{p}^\parallel(s)=1$ ($f_{p}^\parallel(t)=\delta(t)$) and the remaining forcing terms equal to zero. For $0\le\beta\le2\pi$ it has been sufficient to consider only the transverse modes of E-type with $\ell=1,2,3$. In Fig. \ref{fig:LongLong} we show the frequency responses $H^{\parallel \parallel}_{pp} \left( \omega \right)$, $H^{\perp \parallel}_{p'p} \left( \omega \right)$, and the impulse responses $h^{\parallel \parallel}_{pp} \left( t \right)$, $h^{\perp \parallel}_{p'p} \left( t \right)$ for increasing normalized particle size $\beta$: $\beta=\pi/2$ (a-b), $\beta=\pi$ (c-d) and $\beta=2\pi$ (e-f).

For $\beta=\pi/2$ the first peak (from the left) of the amplitude of $H^{\parallel \parallel}_{pp} \left( \omega \right)$, labeled with a black triangle in Fig. \ref{fig:LongLong}(a), is located in proximity of the natural frequency of the electric dipole mode $\mathbf{U}_{m1v}^\parallel$, accordingly to Eq. \ref{eq:PlasmonSmallResonances1}. Nevertheless, a bump is present at higher frequency (labeled with a black circle), due to the radiative self-coupling of the longitudinal dipole mode, which implies the presence of a second peak in the frequency response. In Fig. \ref{fig:LongLong} (b) we show the corresponding impulse response $h^{\parallel \parallel}_{pp} \left( t \right)$. The frequency response $H^{\perp \parallel}_{p'p} \left( \omega \right)$ shown in Fig. \ref{fig:LongLong} (c) and the impulse response $h^{\perp \parallel}_{p'p} \left( t \right)$ shown in Fig. \ref{fig:LongLong} (d) show, respectively, the same behavior of $H^{\parallel \parallel}_{pp} \left( \omega \right)$ and $h^{\parallel \parallel}_{pp} \left( t \right)$, but their amplitudes are roughly an order of magnitude lower.

For $\beta = \pi$ the frequency response undergoes a broadening around the first peak (Fig. \ref{fig:LongLong} (c)), while the previous bump becomes a secondary peak. The contribution of the longitudinal-transverse coupling starts to be significant. The faster decay of the corresponding impulse response $h^{\parallel \parallel}_{pp} \left( t \right)$ (Fig. \ref{fig:LongLong} (d)) reflects the broadening of the amplitude response around the first peak, while the beating subtends the interaction between the poles associated to the first two peaks. Figures \ref{fig:LongLong} (g) and \ref{fig:LongLong} (h) show, respectively, the frequency response $H^{\perp \parallel}_{p'p} \left( \omega \right)$ and the impulse response $h^{\perp \parallel}_{p'p} \left( t \right)$.

For $\beta= 2 \pi$ the second peak of the frequency response becomes dominant (Fig. \ref{fig:LongLong} (e)). The corresponding impulse response oscillates with a frequency close to the frequency position of the second peak Fig. (\ref{fig:LongLong} (f)). Moreover, the number of oscillations is higher than in panel (d), corresponding to a lower decay rate. Figures \ref{fig:LongLong} (k) and \ref{fig:LongLong} (l) show, respectively, the frequency response $H^{\perp \parallel}_{p'p} \left( \omega \right)$ and the impulse response $h^{\perp \parallel}_{p'p} \left( t \right)$. They show the same behaviour and the same order of magnitude of $H^{\parallel \parallel}_{pp} \left( \omega \right)$ and $h^{\parallel \parallel}_{pp} \left( t \right)$, respectively.

Figure \ref{fig:LongLong} shows that in metal particles the frequency response of the fundamental longitudinal mode is dominated by multiple peaks which move as a function of $\beta$, and the impulse response is characterized by multiple harmonics. To further investigate this behavior, we use a rational function approximation of the frequency response evaluated by the vector fitting technique \cite{gustavsen_improving_2006,gustavsen_rational_1999,deschrijver_macromodeling_2008}. We consider the first two dominant poles of $H^{\parallel \parallel}_{pp} \left( \omega \right)$ with $p=\left(m1\nu\right)$, which we denote as pole $1$ and pole $2$. Figure \ref{fig:Vector_Fit} show the real and imaginary parts of both poles as a function of $\beta$, together with the magnitudes of the corresponding residues. The pole $1$ corresponds to the peak labeled with a black triangle in Figures \ref{fig:LongLong} (a), \ref{fig:LongLong} (e) and \ref{fig:LongLong}  (i), while the pole $2$ corresponds to the peak labeled with the black circle. In the small size limit the imaginary part of the pole $1$ tends to the natural frequency of the longitudinal dipole mode $\mathbf{U}_{m1v}^\parallel$ normalized to $\omega_P$. The pole $2$, as we will see, is associated to a transverse electromagnetic standing wave of the particle. As shown in Fig. \ref{fig:Vector_Fit} (a), the imaginary part of both poles decreases. The absolute value of the real part of the pole $1$ increases, reaches a maximum and slowly decreases, while the absolute value of the pole $2$ monotonically decreases. Fig. \ref{fig:Vector_Fit} (c) shows that while in the small size limit the residue of pole $1$ prevails by several orders of magnitude, as the size of the particle increases, the residue of pole $2$ increases and eventually becomes dominant.

We now investigate the responses for $\beta>2\pi$ $( a>2\pi  k_p  =\lambda_P)$, where the longitudinal-transverse coupling is strong. Figure \ref{fig:RadiativeRegime} shows the amplitude responses $|H^{\parallel \parallel}_{pp} \left(\omega \right)|$ (a), $|H^{\perp \parallel}_{p'p} \left(\omega \right)|$ (c), and the impulse responses $h^{\parallel \parallel}_{pp} \left(t \right)$ (b), $h^{\perp \parallel}_{p'p} \left(t \right)$ (d) for $\beta=3\pi$, $\beta=4\pi$ and $\beta=5\pi$. In these cases, we have considered the coupling among the longitudinal dipole mode and the transverse mode of E-type with $\ell=1,2,3,4$. As $\beta$ increases, the first peak in both the amplitude responses (Figures  \ref{fig:RadiativeRegime} (a) and (c)), which are located on the left of $\omega/\omega_P=1$, continues to move toward lower frequencies as in Fig. \ref{fig:LongLong}. The second peak, while moving to the left, remains confined to the immediate right of $\omega/\omega_P=1$. Moreover, it becomes narrower and grows in amplitude. The other minor peaks to the right of $\omega/\omega_P=1$ behave in the same way as $\beta$ increases. All the peaks to the right of $\omega/\omega_P=1$ are associated to the natural frequencies of the standing transverse electromagnetic waves of the particle, which are in cutoff for $\omega<\omega_P$. For $\beta>2\pi$ the impulse responses $h^{\parallel \parallel}_{p'p} \left(\omega \right)$ and $h^{\perp \parallel}_{p'p} \left(\omega \right)$ are dominated by the component associated to the second peak in the amplitude responses, this give arise to beatings (Figures  \ref{fig:RadiativeRegime} (b) and (d)). The decay rate of the impulse responses decreases as the second peak in the amplitude becomes narrower, as $\beta$ increases.

Other scenarios have been investigated: the frequency and impulse response of higher order multipolar longitudinal and transverse modes of E-type; in both scenarios the forcing terms is either longitudinal or transverse (E-type). We found that the frequency response of longitudinal modes of higher order does not qualitatively differ from the one described here. When the forcing terms are transverse (of E-type), the coupling with the longitudinal modes is very significant even in the small size limit, because the longitudinal mode may resonate. The opposite was not true, since the transverse modes are of resonance in metal particles.

\subsection{Frequency and impulse response of the mode amplitudes in a dielectric particle} 

\begin{figure}[ht]
    \centering
    \includegraphics[width=\columnwidth]{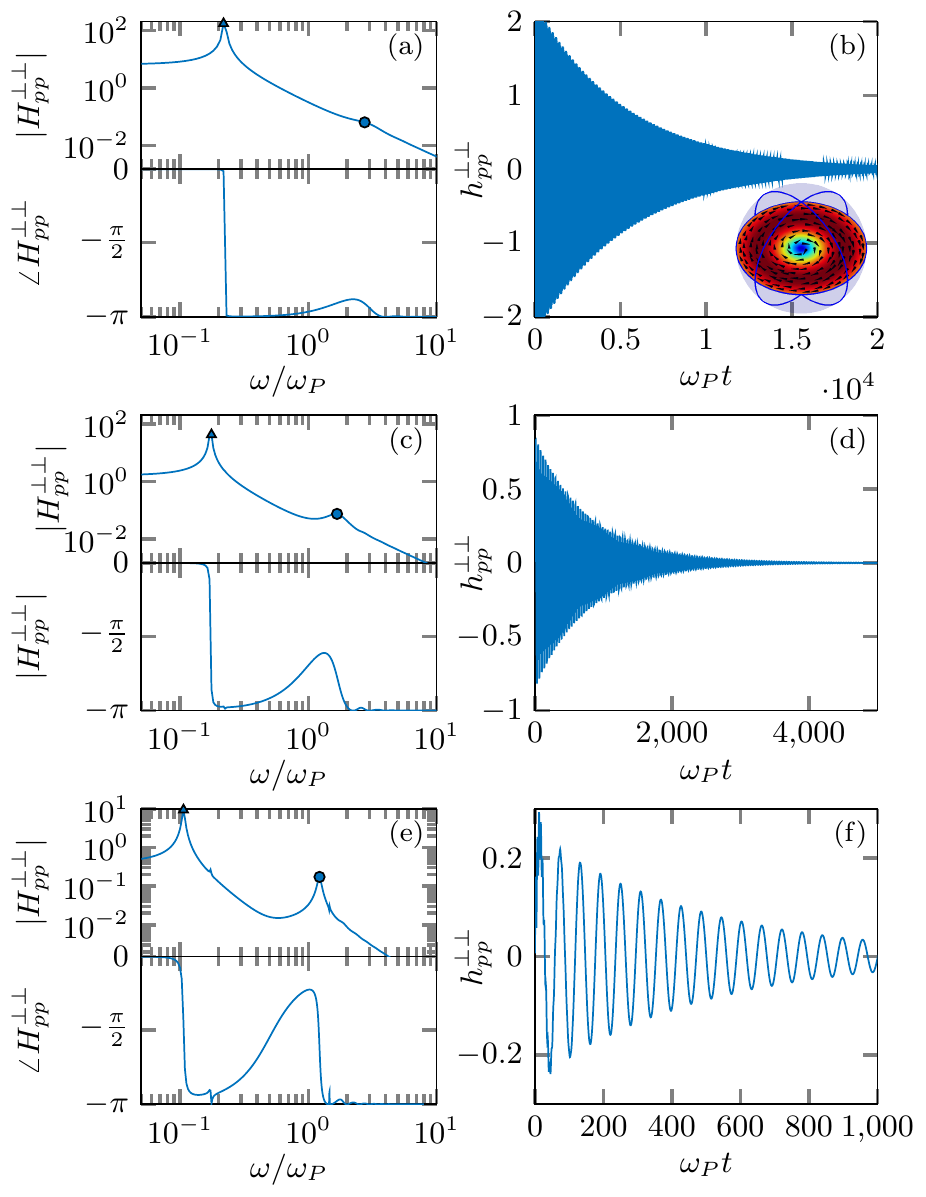}
        \caption{Frequency response $H^{\perp \perp}_{pp}$ (first column) and impulse response $h^{\perp \perp}_{pp}$ (second column) of the transverse magnetic dipole mode amplitude with $p=\left(m111v\right)$ of a lossless dielectric sphere ($\omega_0 = \omega_P/4$, $\Gamma=0$) with $\beta=\pi/2$ (a-b), $\beta=\pi$ (c-d) and $\beta=2\pi$ (e-f) where $\beta=k_Pa$, $a$ is the radius and $k_P=\omega_P/c_0$.}  
    \label{fig:TEresponse}
\end{figure}

\begin{figure}
    \centering
    \includegraphics[width=\columnwidth]{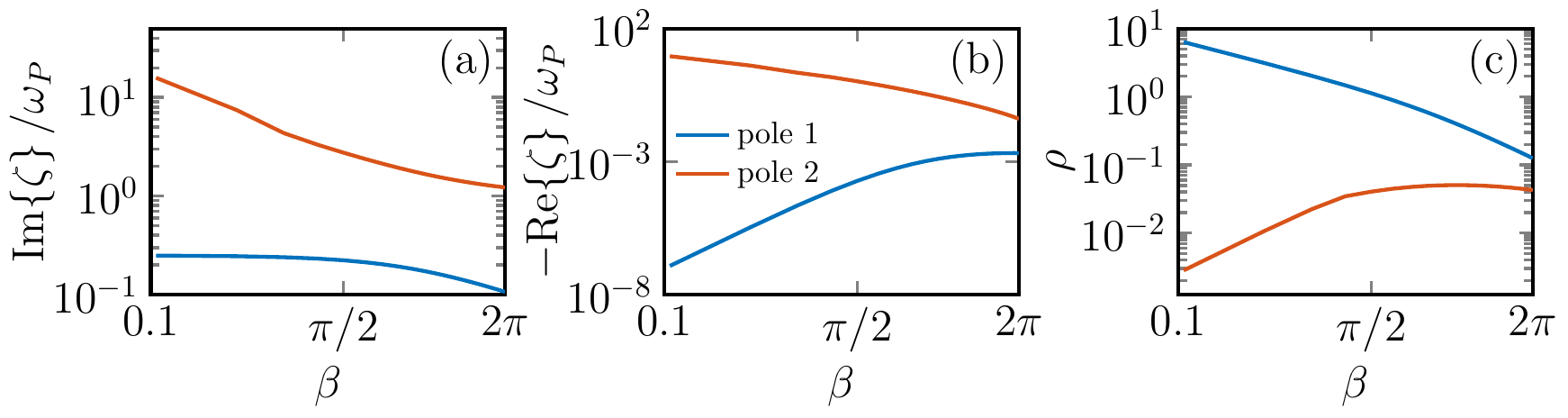}
    \caption{Imaginary (a) and real (b) parts of the first two dominant poles, with the corresponding magnitude of the residues (c) of the frequency response $H^{\perp\perp}_{pp}$of the transverse mode of H-type ($n=1$, $l=1$) of a lossless dielectric sphere ($\omega_0 = \omega_P/4$, $\Gamma=0$) as a function of $\beta= k_Pa$ where $a$ is the radius and $k_P=\omega_P/c_0$. These quantities are obtained by using the vector fitting technique.} 
    \label{fig:VectorFit_TE}
\end{figure}

\begin{figure*}
    \centering
    \includegraphics[width=0.95\textwidth]{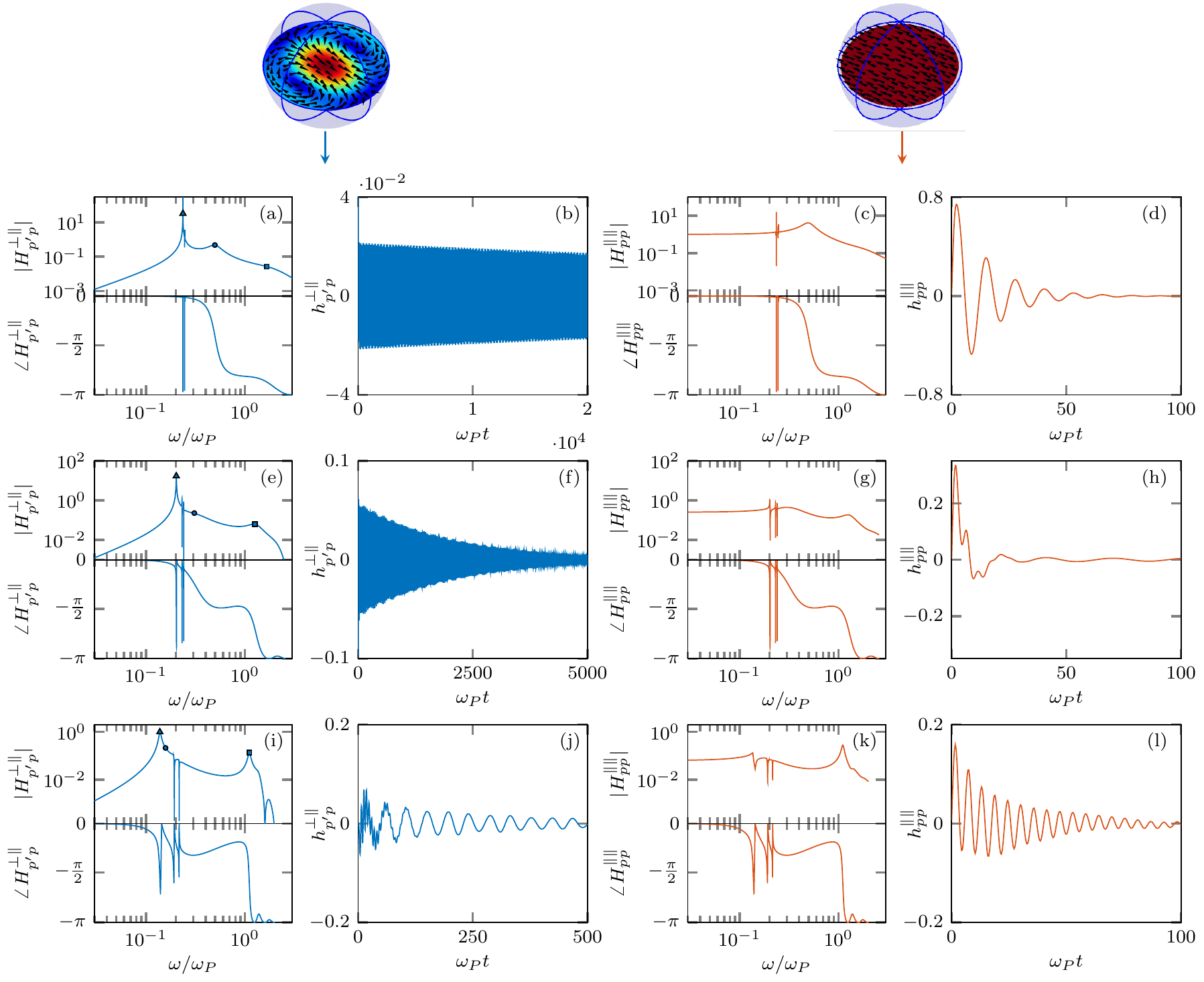}
    \caption{Frequency response $H^{\perp \parallel}_{p'p}$ (first column) and impulse response $h^{\perp \parallel}_{p'p}$ (second column) of the  E-type transverse mode amplitude with  $p=\left(m1\nu\right)$, $p'=\left(m121\nu\right)$ and forcing term $F_p^\parallel=1$; frequency response $H^{\parallel \parallel}_{pp}$ (third column) and impulse response $h^{\parallel \parallel}_{pp}$ (fourth column) of the $p$-longitudinal mode amplitude. We consider a lossless dielectric sphere ($\omega_0 = \omega_P/4$, $\Gamma=0$) with $\beta=\pi/2$, $\beta=\pi$ and $\beta=2\pi$ where $\beta=k_Pa$, $a$ is the radius and $k_P=\omega_P/c_0$.}
    \label{fig:TMresponse}
\end{figure*}

\begin{figure}
    \centering 
    \includegraphics[width=\columnwidth]{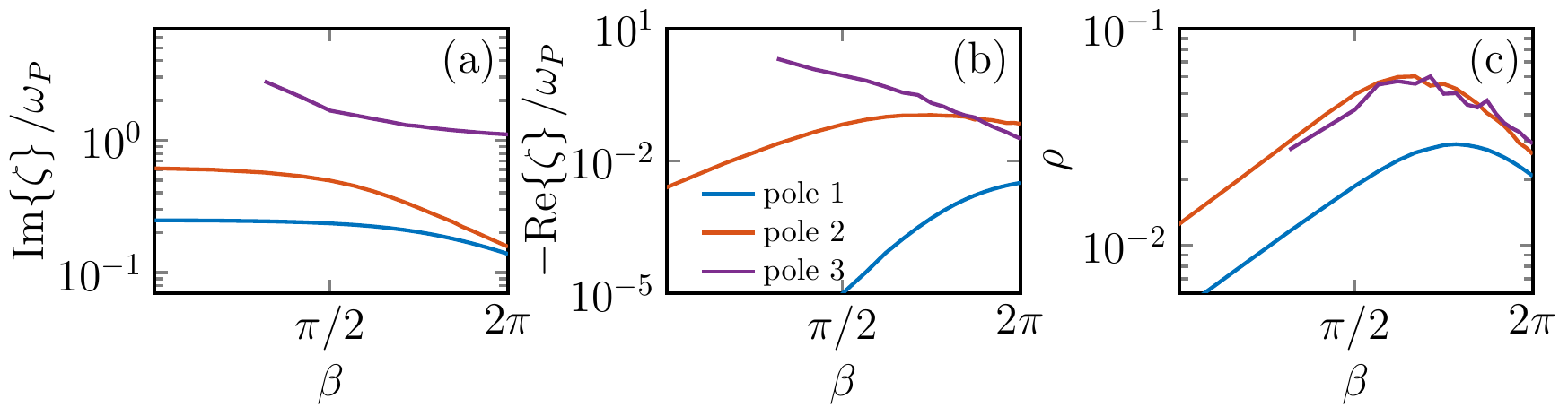}
    \caption{Real and imaginary parts of the three dominant poles with the corresponding magnitude of the residues of frequency response $H^{\perp \parallel}_{p'p}$ of the transverse mode of E-type ($n=1$, $\ell=1$, $s=2$) of a lossless dielectric sphere ($\omega_0 = \omega_P/4$, $\Gamma=0$) as a function of $\beta= k_Pa$ where $a$ is the radius and $k_P=\omega_P/c_0$. These quantities are obtained by using the vector fitting technique.} \label{fig:VectorFit_TM}
\end{figure}

We now investigate the response of a dielectric particle with $\omega_0 = \omega_P/4$, $\Gamma=0$, beyond the small size limit as $\beta=k_P a$ varies in the interval $\left[0,2\pi\right]$; in this case the susceptibility at $\omega=0$ is equal to $16$. For $0\leq \beta \leq 2\pi$ the susceptibility of the Drude-Lorentz model assumes both positive and negative values. 
\subsubsection{Transverse modes of H-type}
We here analyze the response of the amplitudes of the transverse modes of H-type.
We consider the frequency response $H^{\perp \perp}_{pp} \left( \omega \right)$ of the transverse magnetic dipole mode $\mathbf{U}_{p}^\perp$ amplitude, with $p=\left(m111v\right)$, namely $n=1$, $s=1$, and $\ell=1$; in the system of equations \ref{eq:xparallel5}, \ref{eq:xperp5} we have set $F_{p}^\parallel=1$ and the remaining forcing terms equal to zero. This transverse mode is coupled through radiation only to the transverse modes of the same type having the same multipolar order $n$. For $0\leq \beta \leq 2\pi$ it is sufficient to consider only the coupling with the modes with $\ell=2,3$. In Fig. \ref{fig:TEresponse}, we show the frequency response $H^{\perp \perp}_{pp} \left( \omega \right)$ and the corresponding impulse response $h^{\perp \perp}_{pp} \left( t \right)$ for increasing particle size: $\beta = \pi/2$ (a-b), $\beta = \pi$ (b-d), $\beta = 2 \pi$ (e-f). 

For $\beta = \pi/2$ (Fig. \ref{fig:TEresponse} (a)) the peak of the amplitude response (labeled with a black triangle) is located in the neighborhood of the natural frequency of magnetic dipole mode $\mathbf{U}^\perp_p$, accordingly to Eq. \ref{eq:DielectricSmallResonance1}. As for the response of the longitudinal modes in metal particles, a bump arises at higher frequencies (labeled with a black circle), which is associated to a second peak in the frequency response arising from the radiative self-coupling. The corresponding impulse response, shown in Fig. \ref{fig:TEresponse} (b), exhibits a very small decay rate. 

In Fig. \ref{fig:TEresponse} (c) a larger value of $\beta$ is considered, $\beta = \pi$. The first peak undergoes a broadening, while the high-frequency bump becomes a second (minor) peak. The impulse response shown in Fig. \ref{fig:TEresponse} (d) has a higher radiative decay rate compared to the previous case. 

For $\beta = 2 \pi$ the effects of the coupling with the modes with $\ell=2,3$ are important but only at higher frequencies, modifying the response in neighborhood of the second peak (Fig. \ref{fig:TEresponse} (e)). Nevertheless, the first peak is still dominant, and it characterizes the impulse response shown in \ref{fig:TEresponse} (f).

We now use the vector fitting technique \cite{gustavsen_improving_2006,gustavsen_rational_1999,deschrijver_macromodeling_2008} to study the behavior of the two dominant poles of $H^{\perp \perp}_{pp} \left( \omega \right)$ and of the corresponding residues as $\beta$ varies, see Fig. \ref{fig:VectorFit_TE}. In the small size limit, the blue curve associated with the pole $1$, tends to the natural frequency of the transverse magnetic dipole mode $\mathbf{U}_{m1v}^\parallel$. The pole $2$, shown with a red curve, is associated to a transverse electromagnetic standing wave of the particle. In Fig. \ref{fig:TEresponse} (a),(c),(e) the black triangles and circles are associated with the imaginary part of the pole $1$ and $2$, respectively.

The peaks in the amplitude responses, with the exception of the first one, are confined to the right of $\omega/\omega_c$ where $\omega_c=\sqrt{\omega_0^2+\omega_P^2}$ is the cutoff frequency of the medium. These peaks are associated with the natural frequencies of the standing transverse electromagnetic waves of the particle.

\subsubsection{Transverse modes of E-type and longitudinal modes}

We here analyze the response of the longitudinal modes and of the transverse modes of E-type, which are coupled through the radiation field. Due to the longitudinal-transverse coupling, the behavior is richer than in the previous case. 

We study the frequency response $H^{\perp \parallel}_{p'p} \left( \omega \right)$ of the E-type transverse mode $\mathbf{U}_{p'}^\perp$ amplitude with $p'=\left(m121v\right)$ and $p=\left(m1v\right)$, and the frequency response  $H^{\parallel \parallel}_{pp} \left( \omega \right)$ of the longitudinal mode $\mathbf{U}_{p}^\parallel$ amplitude. In the system of equations \ref{eq:xparallel5}, \ref{eq:xperp5}  we have set $F_{p}^\parallel=1$ and the remaining forcing terms equal to zero. In the investigated interval of $\beta$ it is sufficient to consider only the transverse modes of E-type with $\ell=1,2,3$. 

First, we consider the case $\beta=\pi/2$. The amplitude of $H^{\perp \parallel}_{p'p}$, shown in Fig. \ref{fig:TMresponse} (a), has two peaks and one bump. The first peak (labeled with a triangle) is located in proximity of the natural frequency of E-type transverse mode $\mathbf{U}_{p'}^\perp$, accordingly to Eq. \ref{eq:DielectricSmallResonance1}. The second peak (labeled with a circle) is located in proximity of the natural frequency of the longitudinal mode $\mathbf{U}_{p}^\parallel$, accordingly to Eq. \ref{eq:PlasmonSmallResonances1}.  The magnitude of $H^{\parallel \parallel}_{pp}$, shown in Fig. \ref{fig:TMresponse} (c), shows only a peak in proximity of the natural frequency of the longitudinal mode, because a zero cancels the pole associated to the natural frequency of E-type transverse mode. The impulse response $h^{\perp \parallel}_{p'p}$ is dominated by the transverse mode, while the  $h^{\parallel \parallel}_{p'p}$ is dominated by the longitudinal mode. The decay rates of $h^{\perp \parallel}_{p'p}$ is much smaller than the one of $h^{\parallel \parallel}_{p'p}$, consistently with the analysis carried out above in Sect. VII B. Increasing the value of $\beta$, the second peak in the amplitude of $H^{\perp \parallel}_{p'p}$ disappears and the high-frequency bump, denoted by a black square, becomes a peak (Fig. \ref{fig:TMresponse} (e)). The phase also shows several oscillations. This behaviour also occurs for $H^{\parallel \parallel}_{pp}$, and gives arise to the impulse response shown in Fig. 7 (h). 

For $\beta = 2 \pi$ (Fig. \ref{fig:TMresponse} (i)), several interferences among the three dominant poles arise, determining dips in the magnitude and oscillations in the phase of $H^{\perp \parallel}_{p'p}$. The frequency response $H^{\parallel \parallel}_{pp}$ (Fig.  \ref{fig:TMresponse} (k)) shows similar features. Moreover, the amplitude of $H^{\parallel \parallel}_{pp}$ is dominated by the third peak.  The impulse responses $h^{\perp \parallel}_{p'p}$ and $h^{\parallel \parallel}_{p'p}$ shown in Fig. \ref{fig:TMresponse} (j) and (l), respectively, show an initial rapid oscillation dominated by the radiative coupling, which decays very quickly. This is particular intense in the case of $h^{\parallel \parallel}_{p'p}$. After the decay of this frequency component, both impulse responses show a long period of oscillation.

We now use the vector fitting technique \cite{gustavsen_improving_2006,gustavsen_rational_1999,deschrijver_macromodeling_2008} to study the behaviour of the three dominant poles of $H^{\perp \parallel}_{p'p}$ and of the corresponding residues as $\beta$ varies, see Fig. \ref{fig:VectorFit_TM}. In the small size limit, the blue and red curves, associated with the pole $1$ and $2$ respectively, tend to the natural frequency of the transverse mode $\mathbf{U}_{p'}^\perp$ of E-type and of the longitudinal mode $\mathbf{U}_{m1v}^\parallel$. The pole $3$, shown with a red curve, is associated to a transverse electromagnetic standing wave of the particle. In Fig. \ref{fig:TMresponse} (a),(e) and (i), the black triangles, circles and squares are associated with the imaginary part of the pole $1$, $2$, and $3$, respectively.

The peaks in the amplitude responses, except for the first two, are also confined to the right of $\omega/\omega_c$ where $\omega_c=\sqrt{\omega_0^2+\omega_P^2}$ is the cutoff frequency of the medium. These peaks are associated with the natural frequencies of the transverse electromagnetic standing waves of the particle.

\section{Conclusions} 
\label{sec:Conclusion}
We have introduced a novel formulation for the full wave analysis of time evolution of the polarization induced in the electromagnetic scattering by dispersive particles. In the framework of the {\it Hopfield model} for dielectrics, we expand the polarization field in terms of the static {\it longitudinal (electrostatic) and transverse (magnetostatic) modes} of the particle, and the radiation field in terms of the {\it transverse electromagnetic wave modes} of free space. This choice allows us to separate effectively the role of the polarization field from the role of the radiation field, and to analyze their interaction. We use the {\it principle of least action} to determine the equations governing the time evolution of the mode amplitudes. We also introduce the losses of the matter through a linear coupling of the polarization field to a bath of harmonic oscillators with a continuous range of natural frequencies. We then reduce the set of linear integro-differential equations governing the overall system by eliminating the degrees of freedom of the radiation field and of the bath field. The reduced system governs the time evolution of the amplitude of the longitudinal and transverse polarization modes. We study this system in detail and find the principal characteristics of its temporal evolution, with emphasis on the impulse responses. We found that the temporal evolution is strongly influenced by the self- and mutual coupling between the degrees of freedom of the polarization field that are mediated by the radiation field.

The parameter $k_P a$, where $k_P = \omega_ P / c_0$ and $a$ is the radius of the smallest sphere enclosing the particle, plays an essential role. The {\it plasma wavelength} $\lambda_P = 2 \pi / k_P$ appears as the natural characteristic dimension to describe how the matter interacts with the electromagnetic field.

To investigate the role of the coupling, we analyse the system of equations in the Laplace domain.  We found that the coupling between the polarization modes is mediated by the full-wave {\it transverse} dyadic Green function in the vacuum. We also found that the static longitudinal (electrostatic) modes and the static transverse (magnetostatic) modes of the particle are the natural oscillation modes of the polarization in the small size limit $a \ll \lambda_P$, and we provide their natural frequencies in a closed form.

We apply the developed approach to a spherical particle with radius $a$. Its static longitudinal and transverse polarization modes have analytic expressions. The transverse  modes are divided into two subsets: the ones of E-type and the ones of H-type. We determine the semi-analytical expressions of the coupling coefficients governing their interactions, and deduce a set of selection rules: i) each mode interacts with itself; ii) only modes with the same multipolar order may interact; iii) the transverse modes of H-type do not interact with either longitudinal modes or transverse modes of E-type ; iv) longitudinal modes and transverse modes of E-type interact. After validating the proposed approach, we investigate the frequency and impulse responses of metal and dielectric particles for $0\leq a \leq 2.5 \lambda_P$. In metal particles, we analyze the evolution of the electric dipole mode amplitude, which interacts significantly only with a few E-type transverse modes of the same multipolar order. In a dielectric particle, we analyse two scenarios. In the first one, we excite the sphere along a transverse degree of freedom of H-type, which interacts significantly only with few H-type transverse modes of the same multipolar order, and then we follow the evolution of the magnetic dipole mode amplitude. In the second scenario, we excite the dielectric sphere along a longitudinal degree of freedom, then we monitor the evolution of both longitudinal and E-type transverse mode amplitudes. Moreover, in this case the interaction couples significantly a longitudinal mode with few E-type transverse modes. In all investigated cases, the frequency response is dominated by multiple peaks that move as the radius of the particle increases. There are peaks associated to the natural frequencies of the static longitudinal and transverse modes of the particle, and peaks associated to the natural frequencies of the transverse electromagnetic standing waves of the particle. These standing waves exhibit a cut-off frequency greater than $\sqrt{\omega_0^2 + \omega_P^2}$. This behavior is investigated in detail, employing the vector fitting technique to achieve a rational function approximation of the frequency response. The impulse responses reflect the frequency behavior, and it is often characterized by a beating between different frequencies, with different decay times. 

Specifically, we found that in the small size limit, $a \ll \lambda_P$, the principal characteristics of the impulse responses are mainly determined by the radiative self-coupling of the static longitudinal and transverse modes of the particle, which are the natural modes of the polarization. The self-coupling is responsible for the shift of their natural frequencies and for the decay. Instead, the mutual-coupling mainly determine the energy transfer between the static transverse and longitudinal modes, which is a non-conservative process due to the radiation losses. This mutual coupling is weak for small size particles, but it becomes relevant as the size of the particle increases. The choice to expand the polarization in terms of the static longitudinal and transverse modes of the particles turns out to be very appropriate to describe the electromagnetic scattering from dispersive particles with size of the order of $\lambda_P$. Since the longitudinal and transverse modes of the particle are the natural modes of the polarization in the small size limit, the interaction due to the radiation involves only a few longitudinal and transverse modes in particles with $ a \sim \lambda_P$. For $a>\lambda_P$, the role of the transverse electromagnetic standing waves becomes important and higher order transverse polarization modes are excited. 

The proposed approach leads to a general method for the analysis of the temporal evolution (transients and steady states) of the polarization field induced in dispersive particles of any shape. First, the longitudinal and transverse modes of the particle are computed by solving static eigenvalue problems by using standard tools of computational electromagnetism. Then, the interaction integrals in the frequency domain are evaluated by properly addressing the $1/r$ singularity. Eventually, the linear coupled system is solved, and the corresponding solution is antitransformed. Only few static longitudinal and transverse modes are needed to properly describe the response of an object even when its size is larger than the characteristic wavelength associated with the material. Once the polarization field is known, the time evolution of the scattered electromagnetic field can be computed inside the particle by using equations \ref{eq:disp}, \ref{eq:res}, and outside the particle by using the electromagnetic potentials. The separation of the contribution of the matter (polarization) from the radiation leads to a physical description of: i) the electromagnetic field - matter interaction, especially when the near field coupling is dominant; ii) the interaction between the discrete modes of the matter and the continuum of the radiation field, including interference effects. It also offers a framework where the quantization of the electromagnetic field - matter interaction can be carried out for particles of any shape.

\newpage
\appendix


\section{Transverse wave modes for the free space}
\label{sec:A}

In this paper we use both the transverse vector plane waves and the transverse vector spherical waves as a basis for representing $\Ab^\perp$.

\subsubsection{Transverse vector plane wave modes}

The transverse vector plane wave mode is given by
\begin{equation}
\label{eq:planewave}
    \mathbf{f}_q \rp = \frac{1}{\left(2\pi\right)^{3/2}} \eps_{s,\mathbf{k}} e^{i \mathbf{k} \cdot \rb},
\end{equation}
where ${\bf k} \in \mathbb{R}^3$ is the propagation vector,  $\left\{ \eps_{s,\mathbf{k}}\right\}$ are the polarization unit vectors with $\eps_{s,\mathbf{k}} = \eps_{s,-\mathbf{k}}$ and $s=1,2$ (e.g., \cite{cohen-tannoudji_photons_1997}). 
The two polarization vectors are orthogonal among them,  $\eps_{1,\mathbf{k}} \cdot \eps_{2, \mathbf{k}} = 0$, and are both transverse to the propagation vector, $\eps_{1,\mathbf{k}} \cdot \mathbf{k} = \eps_{2, \mathbf{k}} \cdot \mathbf{k}=0$. Indeed, it is
\begin{equation}
\label{eq:FourierI}
    \eps_{1,\mathbf{k}}\otimes\eps_{1,\mathbf{k}}+\eps_{2,\mathbf{k}}\otimes\eps_{2,\mathbf{k}}+\hat{\mathbf{k}}\otimes\hat{\mathbf{k}}=\overleftrightarrow{I}
\end{equation}
where $\overleftrightarrow{I}$ is the identity dyad and $\otimes$ is the dyadic product.
In this case $q$ is a multi-index corresponding to the pair of parameters $\mathbf{k}$ and $s$, $q = \left(\mathbf{k},s \right)$, and $\sum_q \left( \cdot \right)$ denotes  $\displaystyle\sum_{s=1}^2 \int_{\mathbb{R}^3} \dk\,\left( \cdot \right) $. 
Since $ \Ab^\perp$ is real we have $a_q^*$ = $a_{-q}$ where the index $-q$ denotes the pair $(-\mathbf{k},s)$.
The set of functions  $\left\{ \fb_q \right\}$ are orthonormal:
\begin{equation}
    \langle \fb_{q'}, \fb_q  \rangle = \delta_{s',s} \delta \left( \mathbf{k} - \mathbf{k}' \right).
\end{equation}
\subsubsection{Transverse vector spherical wave modes}

To describe the vector spherical wave modes, we need the spherical coordinates. We denote with $\left(r,\theta,\phi\right)$ the spherical coordinates of the point with position vector $\rb$. The transverse vector spherical wave functions $\mathbf{M}_{mn} \left( \rb ; k \right)$, $\mathbf{N}_{mn}(\rb;k)$ are given by (e.g., \cite{bohren2008absorption})

\begin{subequations}
\begin{eqnarray}
    \mathbf{M}_{mn}&=&\nabla\times \left(\rb\,\psi_{mn}\right),\\
    \mathbf{N}_{mn} &=&\frac{1}{k}\nabla\times\nabla\times\left(\rb\,\psi_{mn}\right),
\end{eqnarray}
\end{subequations}
where (generating function)
\begin{equation}
    \psi_{mn}(\rb;k) = \sqrt{\frac{2}{\pi}}{\frac{k}{\sqrt{n(n+1)}}} j_n(kr) Y_n^m (\theta,\phi),
\end{equation}
$0\leq k<\infty$, $n=1,2,...$, $-n\leq m\leq +n$, $ j_n(kr)$ is the spherical Bessel function of order $n$, and $Y_n^m (\theta,\phi)$ is the spherical harmonic of degree $n$ and order $m$.

The vector fields $\mathbf{M}_{mn} (\rb;k)$ and $\mathbf{N}_{mn} (\rb; k')$ are orthogonal in the three dimensional space (namely, $\mathbf{M}_{mn} (\rb;k)\cdot\mathbf{N}_{mn} (\rb;k')=0$), and satisfy the symmetrical relations
\begin{subequations}
\begin{eqnarray}
\label{eq:spherwave}
    \mathbf{N}_{mn} =\frac{1}{k}\nabla\times\mathbf{M}_{mn},\\
    \mathbf{M}_{mn} =\frac{1}{k}\nabla\times\mathbf{N}_{mn},
\end{eqnarray}
\end{subequations}
and 
\begin{subequations}
\begin{eqnarray}
\label{eq:spherwave}
    \mathbf{M}_{-mn}=\mathbf{M}_{mn}^*,\\
   \mathbf{N}_{-mn}=\mathbf{N}_{mn}^*.
\end{eqnarray}
\end{subequations} 
They are also orthonormal in the Hilbert space:
\begin{subequations}
\begin{eqnarray}
    \langle \mathbf{M}_{m'n'} (k'), \mathbf{M}_{mn} (k) \rangle_{V_\infty}  = \delta_{m',m} \delta_{n',n}\delta \left(k - k'\right), \qquad\\
    \langle \mathbf{N}_{m'n'} (k'), \mathbf{N}_{mn} (k) \rangle_{V_\infty}  = \delta_{m',m} \delta_{n',n}\delta \left(k - k'\right). \qquad
\end{eqnarray}
\end{subequations}
For the transverse vector spherical wave functions, the label $q$ is a multi-index constituted by the set of parameters $(m,n,s,k)$ with $s=1$ if $\mathbf{f}_q =\mathbf{M}_{mn}$ (H-type modes), and $s=2$ if $\mathbf{f}_q =\mathbf{N}_{mn}$ (E-type modes), \cite{bohren2008absorption}.
The symbol $\sum_q \left( \cdot \right)$ denotes  $\displaystyle\sum_{s=1}^2\sum_{m=-n}^{n}\sum_{n=1}^\infty \int_{0}^\infty dk\,\left( \cdot \right)$. Since $ \Ab^\perp$ is real we have $a_q^*$ = $a_{-q}$ where now the label $-q$ denotes the set $(-m,n,s,k)$. To carry out the calculations, it is convenient to express the vector spherical wave functions in terms of the vector spherical harmonics, see Appendix B.

\section{Vector spherical harmonics}

We now use a spherical coordinate system. The spherical coordinates of the point with position vector $\rb$ are $(r,\theta,\phi)$ (with $0 \le r < \infty$, $0\le \theta < \pi$ and  $0\le \phi<2\pi$). The basis for the three-dimensional vector space is the set $(\hat{\rb}, \hat{\bm{\theta}},\hat{\bm{\phi}})$, where $\hat{\rb}$ is the radial unit vector, $\hat{\bm{\theta}}$ is the polar unit vector, and $\hat{\bm{\phi}}$ is the azimuthal unit vector.

The spherical harmonic $Y_n^m \left( \theta,\phi\right)$ of degree $n$ and order $m$, with $n=0,1,2,..$ and $-n\le m \le+n$, is given by
\begin{equation}
     Y_n^m(\theta,\phi)=C_{mn} P_n^{|m|}(cos\theta)e^{im\phi}
\end{equation}
where $P_n^m(cos\theta)$ is the associated Legendre polynomial of degree $n$ and order $m$, and $C_{mn}$ is a normalization coefficient. The spherical harmonics are orthogonal in the Hilbert space. We normalize them in such a way that
\begin{equation}
    \int_{}\,|Y_n^m(\theta,\phi)|^2\,d\Omega = 1,
\end{equation}
where $\int_{}(\cdot)d\Omega = \int_{0}^{\pi}d\theta\, sin\theta\int_{0}^{2\pi}d\phi\,(\cdot)$. The normalization coefficient $C_{mn}$ is equal to
\begin{equation}
    C_{mn} = \sqrt{\frac{2n+1}{4\pi}\frac{(n-m)!}{(n+m)!}}.
\end{equation}

The vector spherical harmonics $\mathbf{Y}_n^m(\theta,\phi)$, $\mathbf{X}_n^m(\theta,\phi)$ and $\mathbf{W}_n^m(\theta,\phi)$ are defined as
\begin{subequations}
\begin{eqnarray}
    \mathbf{Y}_n^m&=&\hat\rb Y_n^m,\\
    \mathbf{X}_n^m&=&\frac{1}{\sqrt{n(n+1)}}\nabla Y_n^m \times \rb,\\
    \mathbf{W}_n^m&=&\hat\rb\times \mathbf{X}_n^m=\frac{1}{\sqrt{n(n+1)}} r\,\nabla Y_n^m.
\end{eqnarray}
\end{subequations}
They are orthogonal in the three dimensional space, namely, $\mathbf{Y}_n^m(\theta,\phi)\cdot \mathbf{X}_n^m(\theta,\phi)=0, 
\mathbf{X}_n^m(\theta,\phi)\cdot \mathbf{W}_n^m(\theta,\phi)=0,  \mathbf{W}_n^m(\theta,\phi)\cdot \mathbf{X}_n^m(\theta,\phi)=0$), and are orthonormal in the Hilbert space,
\begin{subequations}
\begin{eqnarray}
    \int_{}\mathbf{Y}_{n'}^{m'\,*} \cdot\mathbf{Y}_{mn} d\Omega &=& \delta_{m'm}\delta_{n'n},\\
    \int_{}\mathbf{X}_{n'}^{m'\,*} \cdot\mathbf{X}_{mn} d\Omega &=& \delta_{m'm}\delta_{n'n},\\
    \int_{}\mathbf{W}_{n'}^{m'\,*}\cdot\mathbf{W}_{mn} d\Omega &=& \delta_{m'm}\delta_{n'n}.
\end{eqnarray}
\end{subequations}

The spherical vector wave functions $\mathbf{M}_{mn} \left( \rb ; k \right)$, $\mathbf{N}_{mn}(\rb;k)$ can be expressed in terms of the vector spherical harmonics as
\begin{eqnarray}
    \mathbf{M}_{mn} &=& \sqrt{\frac{2}{\pi}}\,k\,j_n(kr)\mathbf{X}_n^m(\theta,\phi),
\end{eqnarray}
\begin{eqnarray}
\nonumber
    \mathbf{N}_{mn} &= \sqrt{\frac{2}{\pi}}\frac{1}{r}\left[\sqrt{n(n+1)}j_n(kr)\mathbf{Y}_n^m (\theta,\phi) \right.
   \\&+ \left. \left(rj_n\left(kr\right)\right)'\,\mathbf{W}_n^m \left(\theta,\phi\right)\right]
\end{eqnarray}
where $j_n(kr)$ is the first kind spherical Bessel function of order $n$; we have denoted the first order derivative of $rj_n(kr)$ with respect to the radial coordinate with $(rj_n)'$. By using the property
\begin{equation}
    \int_{0}^\infty j_n(k'r)j_n(kr)r^2dr = \frac{\pi}{2}\frac{\delta(k'- k)}{k^2},
\end{equation}
and the orthogonality properties of the vector spherical harmonics it follows immediately that the vector spherical wave functions are orthonormal in the Hilbert space.
\section{Longitudinal and transverse modes of a sphere}
\label{sec:B}
The longitudinal and the transverse modes of a sphere are expressible in terms of spherical harmonics and spherical Bessel functions, see Appendix B. We use a system of spherical coordinates with the origin at the centre of the sphere, and indicate with $a$ the sphere radius.
\subsection{Longitudinal modes}
The longitudinal modes $\left\{ \mathbf{U}_p^\parallel \right\}$ of the sphere are characterized by three indexes $p=(m,n,v)$, where $n=1,2,...$, $0\leq m \leq n$, $v=e$ (even modes) and $o$ (odd modes). We have 
\begin{equation}
\mathbf{U}_{ m n\,\substack{ e\\ o} }^\parallel(\rb) =  \frac{1}{\sqrt{c_{m n}^\parallel}}\left( \begin{array}{c} 1  \\ -i  \end{array}\right) \left[  \mathbf{U}_{m n}^{\parallel} \left( \rb \right)  \pm\mathbf{U}^{\parallel}_{-m n}(\rb) \right]
\label{eq:EQSmode}
\end{equation}
where
\begin{equation}
    \mathbf{U}_{m' n}^\parallel (\rb)= r^{n-1} \left[n\mathbf{Y}_n^{m'} \left(\theta,\phi\right)+\sqrt{n(n+1)}\mathbf{W}_n^{m'} \left( \theta,\phi \right) \right]
\end{equation}
and 
\begin{equation}
    c_{m n}^{\parallel}=2na^{2n+1}(1+\delta_{m 0}).
\end{equation}
The normalization constant $c_{m n}^\parallel$ has been chosen in such a way $\left\| \mathbf{U}_{m n v}^\parallel \right\|=1$. 
The eigenvalue $\lambda_n$ associated to the mode $\mathbf{U}_{m n v}^\parallel(\rb)$ is given by $ \lambda_n= (2n+1)/n$. It does not depend on the indexes $v$ and $m$.

\subsection{Transverse modes}

There are two kinds of transverse modes $\left\{ \mathbf{U}_p^\perp \right\}$: the vector fields that are orthogonal to the radial direction $\hat{\rb}$ (H-type transverse modes), and the the vector fields that have a radial component different from zero (E-type transverse modes). The transverse modes $\left\{ \mathbf{U}_p^\perp \right\}$ of the sphere are characterized by five indexes, $p=(m,n,s,\ell,v)$, where $0\leq m \leq n$, $n=1,2,...$, $s=1$ for the H-type and $s=2$ for the E-type modes, $\ell=1,2,...$, $v=e$ (even modes) and $v=o$ (odd modes). We have
\begin{equation}
\begin{aligned}
\mathbf{U}_{m \,n\, s\,\ell\,\substack{ e\\ o}}^\perp \left(\rb\right) &=  \frac{1}{\sqrt{c_{m n s l}^\perp}} \left( \begin{array}{c} 1  \\ -i  \end{array}\right) \left[  \mathbf{U}_{m n s \ell}^\perp \left(\rb\right)  \pm \mathbf{U}_{-m n s \ell}^\perp(\rb)\right],
\end{aligned}
\label{eq:EQSmode}
\end{equation}
where for the H-type modes
\begin{equation}
    \mathbf{U}_{m'\, n\, 1\, \ell}^\perp(\rb)=j_n \left(z_{(n-1),\ell} \, r \right/a)\mathbf{X}_n^{m'}(\theta,\phi),
\end{equation}
and for the E-type modes
\begin{eqnarray}
\;\nonumber
\mathbf{U}_{m'\, n\,2\,\ell}^\perp(\rb)&=&\frac{1}{\kappa_{n,\ell}\,r} \{
\\
 \;\nonumber
&&\sqrt{n(n+1)}\,j_n(z_{n,\ell} \, r/a) \mathbf{Y}_n^{m'}(\theta,\phi)
    \\
    &+&\frac{d}{dr}[rj_n(z_{n,\ell} \, r/a)]\,\mathbf{W}_n^{m'}(\theta,\phi)\};
\end{eqnarray}
here $z_{m,\ell}$ is the $\ell$-th zero of the spherical Bessel function of order $m$, $j_m$.The normalization constant $c_{m n s \ell}$ has been chosen in such a way $\left\| \mathbf{U}_{m n s \ell v}^\perp \right\|=1$,
\begin{equation}
    c_{m\, n\, s\, \ell}^\perp=a^3(1+\delta_{m0})d_{n s \ell}^\perp,
\end{equation}
where for the H-type modes
\begin{equation}
    d_{n\,1\, \ell}^\perp=j_n^2(z_{(n-1),\ell}),
\end{equation}
and for the E-type modes
\begin{equation}
    d_{n \,2\, \ell}^\perp=\frac{1}{2n+1}[(n+1)j_{n-1}^2(z_{n,\ell})+nj_{n+1}^2(z_{n,\ell})].
\end{equation}
The eigenvalue $\kappa_{n,\ell}$ associated to the transverse mode of H-type $\mathbf{U}_{m\, n\, 1\, l\, v}^\perp$ is given by $\kappa_{n,\ell}=(z_{(n-1),\ell}/a)^2$, and to the transverse mode of E-type $\mathbf{U}_{m\, n \,2\, l\, v}^\perp$ is given by $\kappa_{n,\ell}=(z_{n,\ell}/a)^2$. As for the longitudinal modes, the eigenvalues do not depend on the indexes $v$ and $m$.

\section{Expressions of the coefficients $S_{pp'}^{a\,b}(s)$ at $s=i\omega + \epsilon$ where $\epsilon \downarrow 0$ for a sphere}
\label{sec:Coefficients}
In this Appendix, we first give the expressions of the scalar products $\langle \fb_{q},\Ua_{p}\rangle$ for the longitudinal and transverse modes of a sphere with radius $a$, then, we calculate the expressions of the coefficients $S_{pp'}^{a\,b}(s)$, and at the end we give their expressions evaluated for $s=i\omega + \epsilon$ where $\epsilon \downarrow 0$. It is convenient to use as basis for the transverse component of the vector potential the transverse vector spherical wave functions. The function $\fb_{q}$ is characterized by four indexes: $q=(\mt,\nt,\st,k)$ where $-\nt\leq \mt \leq +\nt$, $\nt=1,2,...$, $\st=1$ for the H-type modes, $\st=2$ for the E-type modes, and $0\leq k < \infty$.

\subsection{Expression of the scalar product $\langle \fb_{\mt\,\nt\,\st\, k},\mathbf{U}^\parallel_{m\, n \, v}\rangle$}

The longitudinal modes are orthogonal to the vector spherical harmonic $\mathbf{X}_n^m$, therefore $\langle \fb_{\mt\,\nt\,\st\,k},\mathbf{U}^\parallel_{m\, n \,v}\rangle=0$ for $\st= 1$. By using the properties of the vector spherical harmonics and of the spherical Bessel functions we obtain
\begin{equation}
    \langle\fb_{\mt\,\nt\,\st\,k},\mathbf{U}^\parallel_{m\,n\, v}\rangle=\delta_{|\mt|m}\,\delta_{\nt n}\delta_{\st 2}\,\,W_{m\, n\,v}^\parallel(ka),
\end{equation}
where
\begin{equation}
    W_{m n v}^\parallel(ka)=(1+\delta_{m0})\sqrt{\frac{2n(n+1)}{\pi c_{ m n}^\parallel}}a^{n+1}w_{m n v}^\parallel(ka),
\end{equation}
\begin{equation}
    w_{m\,n\,e}^\parallel(ka)=j_n(ka),
\end{equation}
and
\begin{equation}
   w_{m \, n\,o}^\parallel(ka)=-i\, \text{sgn} \left(m\right) \, w_{m\,n\,e}^\parallel(ka).
\end{equation}
\subsection{Expression of the scalar product $\langle \fb_{\mt\, \nt\,\st\,  k},\mathbf{U}^\perp_{m\,n\,s\,\ell\,v}\rangle$}

The H-type transverse modes are orthogonal to the E-type vector spherical waves, therefore $\langle \fb_{\mt\, \nt\,\st\,  k},\mathbf{U}^\perp_{m\,n\,s\,\ell\,v}\rangle=0$ for $\st\neq s$. By using the properties of the vector spherical harmonics and of the spherical Bessel functions, we obtain
\begin{equation}
    \langle\fb_{ \mt\,\nt\,\st\, k},\mathbf{U}^\perp_{m\,n\,s\,\ell\,v}\rangle=\delta_{|m|m'}\,\delta_{nn'}\,\delta_{ss'}\,W_{mnslv}^\perp(ka),
\end{equation}
where
\begin{equation}
    W_{mnslv}^\perp(ka)=(1+\delta_{m0})\sqrt{\frac{2}{\pi c_{mnsl}^\perp}}a^{2}w_{mnslv}^\perp(ka),
\end{equation}
\begin{equation}
    w_{m\,n\,s=1\,\ell\,v=e}^\perp(ka)=j_n(z_{n-1,\ell})\,\frac{(ka)^2}{z_{n-1,\ell}^2 - (ka)^2}\,j_{n-1}(ka),
\end{equation}
\begin{equation}
    w_{m\,n\,s=2\,\ell\,v=e}^\perp(ka)=\frac{1}{2n+1}\frac{ka}{z_{n,\ell}^2 - (ka)^2}u_{nl}(ka),
\end{equation}
\begin{eqnarray}
\;\nonumber
u_{nl}(ka)&=&(n+1) \left[j_{n-1}(z_{n,\ell}) \, ka \, j_{n-2} \left( ka \right) \right.
\\&-&\;\nonumber
\left. z_{n,\ell}\,j_{n-2} \left(z_{n,\ell}\right)\,j_{n-1}(ka) \right]
\\&+&  n j_{n+1} \left(z_{n,\ell}\right)\,(ka)\,j_{n}(ka),
\end{eqnarray}
and
\begin{equation}
   w_{m\,n\,s\,\ell\,v=o}^\perp(ka)=-i\,sign(m)w_{m\,n\,s\,\ell\,v=e}^\perp(ka).
\end{equation}
\subsection{Expressions of $S_{p\,p'}^{a\,b}(s)$}

Now we can evaluate $S_{p\,p'}^{a\,b}(s)$ starting from the definition \ref{eq:ExpS}, and remembering that $\sum_q \left( \cdot \right)$ denotes  $\displaystyle\sum_{\st =1}^2\sum_{\mt =-\nt}^{\nt}\sum_{\nt =1}^\infty \int_{0}^\infty \left( \cdot \right) dk$. 

The expression of $S_{p\,p'}^{\parallel\,\parallel}(s)$, with $p=(m,n,v)$ and $p'=(m',n', v')$, is given by
\begin{equation}
    S_{pp'}^{\parallel\,\parallel}(s)=\delta_{mm'}\delta_{nn'}\,\delta_{vv'}\,\frac{2(n+1)}{\pi}\Sigma_{mnv}^\parallel(s)
\end{equation}
where
\begin{equation}
    \Sigma_{mnv}^\parallel(s)=\int_{0}^\infty\frac{s}{s^2+\omega_c^2z^2}|w_{mnv}^{\parallel}(z)|^2dz
\end{equation}
and $\omega_c=c_0/a$.

The expression of $S_{pp'}^{\perp\,\perp}(s)$ with $p=(m,n,s,\ell,v)$ and $p'=(m',n',s',\ell',v')$ is given by:
\begin{equation}
    S_{pp'}^{\perp\,\perp}(s)=\delta_{mm'}\delta_{nn'}\,\delta_{ss'}\delta_{vv'}\frac{4}{\pi\sqrt{d_{nsl}d_{nsl'}}}\Sigma_{mns\,ll'\,v}^\perp(s)
\end{equation}
where
\begin{equation}
    \Sigma_{mns\,ll'\,v}^\perp(s)=\int_{0}^\infty\frac{s}{s^2+\omega_c^2z^2}w_{mnslv}^{\perp*}(z)w_{mnsl'v}^\perp(z)dz.
\end{equation}

Now we consider $S_{pp'}^{\perp\,\parallel}(s)$ with $p=(m,n,s,\ell,v)$ and $p'=(m',n',v')$. We obtain:
\begin{equation}
    S_{pp'}^{\perp\,\parallel}(s)=\delta_{mm'}\delta_{nn'}\,\delta_{s2}\delta_{vv'}\,\frac{2}{\pi}\,\sqrt{\frac{2(n+1)}{d_{n2l}^ \perp}}\Sigma_{mnlv}^{\perp\parallel}(s)
\end{equation}
where
\begin{equation}
  \Sigma_{mnlv}^{\perp\parallel}(s)=\int_{0}^\infty\frac{s}{s^2+\omega_c^2z^2}w_{m\,n\,2\,\ell\,v}^{\perp*}(z)w_{m\,n\,v}^\parallel(z)dz.
\end{equation}
At the end, we have $\Sigma_{p' p}^{\parallel\,\perp}(s)= \Sigma_{pp'}^{\perp\,\parallel\,}(s)$.

\subsection{Expressions of $\Sigma_{mnv}^\parallel(s)$, $\Sigma_{mnsll'v}^\perp(s)$ and $\Sigma_{mnlv}^{\perp\parallel}(s)$ at $s=i\omega + \epsilon$ where $\epsilon \downarrow 0$}
The expressions of $\Sigma_{mnv}^\parallel(s)$, $\Sigma_{mnsll'v}^\perp(s)$ and $\Sigma_{mnlv}^{\perp\parallel}(s)$ are of the type
\begin{equation}
  \Sigma(s)=\int_{0}^\infty\frac{s}{s^2+\omega_c^2z^2}f(z)dz
\end{equation}
where $f(z)$ is a regular function given by bilinear forms of spherical Bessel functions. We need to evaluate $|\Sigma(s)$ for $s=i\omega + \epsilon$ where $\epsilon \downarrow 0$. By applying the partial fraction decomposition, we obtain
\begin{equation}
  \frac{s}{s^2+\omega_c^2z^2}= \frac{1}{2} \left( \frac{1}{s+i\omega_cz}+\frac{1}{s-i\omega_cz} \right),
\end{equation}
therefore
\begin{eqnarray}
\;\nonumber
  \Sigma(s=i\omega+\epsilon)&=&-\frac{i}{2\omega_c}\int_{0}^\infty \frac{1}{\omega/\omega_c+z -i\epsilon}f(z)dz
  \\
  &-&\frac{i}{2\omega_c}\int_{0}^\infty \frac{1}{\omega/\omega_c-z-i\epsilon}f(z)dz.\qquad
\end{eqnarray}
By using the relation
\begin{equation}
  \frac{1}{x-i\epsilon}= i\pi\delta(x) + \mathcal{P}\frac{1}{x},
\end{equation}
where $\mathcal{P}$ denotes the Cauchy principal value, we obtain for $\Sigma(\omega)\equiv \Sigma(s=i\omega+\epsilon)$ the following expression
\begin{equation}
  \Sigma(\omega)=\frac{1}{\omega_c} \frac{\pi}{2} \left[ f(|\omega|/\omega_c)
  +i \frac{\omega}{\omega_c} P_f(\omega/\omega_c) \right],
\end{equation}
where
\begin{equation}
  P_f(\omega/\omega_c)=\frac{2}{\pi}\mathcal{P}\int_{0}^\infty \frac{f(z)}{z^2-(\omega/\omega_c)^2}dz.\qquad
\end{equation}

\subsection{On the evaluation of $w_{m\,n\,s\,\ell\,v}^\perp(z)$}

The expression of $w_{m\,n\,s=1\,\ell\,v}^\perp(z)$ contains the function $\frac{j_{n-1}(z)}{z_{n-1,l}^2 - z^2}$ where $z_{n-1,l}$ is the $l-th$ zero of the spherical Bessel function $j_{n-1}(z)$. It gives $0/0$ for $z=z_{n-1,l}$. This function is well defined at $z=z_{n-1,l}$, its value can be evaluated by using the H$\hat{o}$pital's rule. The first derivative of $j_{n}(z)$ is given by
\begin{equation}
  \frac{dj_{n}}{dz}=\frac{n}{z}j_{n}-j_{n+1}.
\end{equation}
By applying the H$\hat{o}$pital's rule, we obtain:
\begin{equation}
 \lim_{z \to z_{n-1,l}}\left[\frac{j_{n-1}(z)}{z_{n-1,l}^2 - z^2}\right]=\frac{j_{n}(z_{n-1,l})}{2z_{n-1,l}}.
\end{equation}
We proceed in the same way to evaluate $\lim_{z \to z_{n,l}} w_{m\,n\,s=2\,\ell\,v}^\perp(z)$.

\section{Small size limit}
\label{sec:SmallParticle}

In this appendix we give some asymptotic expansions to study the behavior of small size particles.

\subsection{Natural modes}

We now analyze the natural modes in the limit $\beta\rightarrow 0$ where $\beta=k_Pa$ and $k_P=\omega_P/c_0$. In this limit, the system of homogeneous equations \ref{eq:xparallelom1} and \ref{eq:xperpom1} reduces to
\begin{eqnarray}
\label{eq:xparallel8}
\nonumber
\left[\frac{\omega_P^2}{\chi(\zeta)}+\Omega_p^2 \right]{Z}_p^{\parallel}+\beta^2 \zeta^2 \sum_{p''} \left(R_{pp''}^{\parallel \, \parallel} Z_{p''}^\parallel
+R_{pp''}^{\parallel\perp}Z_{p''}^\perp\right)=
\\
\mathcal{O}(\beta^3), \qquad
\end{eqnarray}
\begin{align}
\label{eq:xperp8}
\nonumber
\left[\frac{\omega_P^2}{\chi(\zeta)}+\beta^2 \frac{\zeta^2}{a^2\kappa_p} \right]{Z}_{p'}^{\perp}&+
\\
\nonumber
\beta^2 \zeta^2\sum_{p''} \left[ R_{p'p''}^{\perp\parallel} Z_{p''}^\parallel+
\beta^2 \left(\frac{\zeta}{\omega_P}\right)^2 Z_{p'p''}^{\perp\perp} X_{p''}^\perp \right]= 
\\
=\mathcal{O}\left(\beta^5\right),
\end{align}
where
\begin{align}
    \label{eq:SigmaLL}
    R_{pp'}^{\perp\,\perp} &=  \frac{1}{2}  \frac{1}{4 \pi a^4} \int_V \dV \int_V \dV'  \, \mathbf{U}_p^\perp \rp \cdot \mathbf{U}_{p'}^\perp \rpp \left| \rb -\rb'\right|, \\
    \label{eq:SigmaLT}
   R_{pp'}^{\parallel\,\perp} &= \frac{1}{4 \pi a^2} \int_V \dV \int_V \dV'  \,   \frac{ \mathbf{U}_p^\parallel \rp \cdot \mathbf{U}_{p'}^\perp \rpp}{\left| \rb -\rb' \right|}, \\
    \notag
    R_{pp'}^{\parallel\,\parallel} &= \frac{1}{4 \pi a^2} \int_V \dV \int_V \dV' \, \frac{\mathbf{U}_p^\parallel \rp \cdot   \mathbf{U}_{p'}^\parallel \rpp}{\left| \rb -\rb' \right|}, \\
    \label{eq:SigmaTT}
    & + \frac{1}{2} \frac{1}{4 \pi a^2} \oint_{S} \dS \oint_{S} \dS'
    \mathbf{U}^\parallel_{p} \rp \cdot \n {\left| \mathbf{r} - \mathbf{r}' \right|} \mathbf{U}^\parallel_{p'} \rpp \cdot \n',
\end{align}
The quantities $R_{pp'}^{\perp\,\perp}$, $R_{pp'}^{\parallel\,\perp}$, $R_{pp'}^{\parallel\,\parallel}$ and $1/(a^2\kappa_p)$ do not depend on the size of the particle $a$ and on the complex variable $\zeta$, they only depend on the particle shape. Equations \ref{eq:xparallel8} and \ref{eq:xperp8} have been obtained starting from the expression \ref{eq:Sint}, using the identities \ref{eq:IntegralIdentityI}-\ref{eq:IntegralIdentityII}, and the asymptotic expression \ref{eq:tGreenAsym} of the transverse green function. 

For $\beta=0$ the equations of the system \ref{eq:xparallel8} and \ref{eq:xperp8} decouple and the natural modes of polarization are the longitudinal and transverse modes of the particle. The natural frequencies of the $p-$longitudinal polarization modes are solutions of equation
\begin{eqnarray}
 \frac{\omega_P^2}{\chi(\zeta)} +\Omega_p^2+ \beta^2 R_{pp}^{\parallel \parallel} \zeta^2  = \mathcal{O} \left(\beta^3\right).
 \label{eq:smallsizelong}
\end{eqnarray}
 This equation is obtained by solving perturbatively the system of equations \ref{eq:xparallel8} and \ref{eq:xperp8} for small values of $\beta$ in the neighborhood of $\beta=0$. The term $R_{pp}^{\parallel \parallel}$ describes the effects of the self-coupling of the $p-$longitudinal polarization mode. Equation \ref{eq:smallsizelong} yields the same results obtained in the context of volume integral equation formulations of Maxwell's Equations  \cite{forestiere_resonance_2020}.
Similar steps are followed to derive the natural frequencies of the transverse polarization modes. The natural frequencies of the transverse $p-$polarization mode are solution of the equation
\begin{equation}
    \frac{\omega_P^2}{\chi(\zeta)} + \beta^2\frac{\zeta^2}{a^2\kappa_p}  + \beta^4  \frac{D_p^\perp\zeta^4}{\omega_P^2} = \mathcal{O} (\beta^5) \qquad
    \label{eq:smallsizetrans}
\end{equation}
where 
\begin{equation}
D_p^\perp=R_{pp}^{\perp \perp}  - \sum_{p'}  \lambda_{p'} \left(R_{pp'}^{\perp \, \parallel} \right)^2.
\end{equation}
It has been obtained solving perturbatively the system of equations \ref{eq:xparallel8} and \ref{eq:xperp8} for small values of $\beta$ in the neighborhood of $\beta=0$. The term $R_{pp}^{\perp \perp}$ describes the self-coupling of the $p$- transverse polarization mode. The term $R_{pp'}^{\perp \, \parallel}$ arises from the coupling between the $p$- transverse and the $p'$-th longitudinal polarization modes. The mutual interaction term $R_{pp'}^{\perp \, \parallel}$ is equal to zero if the normal component to $\partial V$ of the vector field
\begin{equation}
    \int_V \dV' \, \frac{\mathbf{U}_{p'}^\perp \rpp}{\left| \rb -\rb' \right|}
\end{equation}
is equal to zero. Equation \ref{eq:smallsizetrans} yields the same results obtained in the context of volume integral equation formulations of Maxwell's equations  \cite{forestiere_resonance_2020}.

The dissipation due to the radiation losses appears with the first odd power of $\zeta$ that have been disregarded in the system of equations \ref{eq:smallsizelong} and \ref{eq:smallsizetrans}. They have been considered in \cite{forestiere_resonance_2020}. In Appendix \ref{sec:Asymptotic} we give the asymptotic expressions of the coefficient $S_{pp'}^{a\,b}(\zeta)$ for the longitudinal and transverse modes of a spherical particle that also take into account the radiation losses.

\subsection{Integral Identities and Asymptotic Expression for the Transverse Dyadic Green Function}

For obtaining the small size limit $\beta\rightarrow0$ introduced before we use the integral identities
\begin{multline}
    \int_V \dV \int_V \dV' \, \Ua_{p} \rp  \frac{ \left( \mathbf{r} - \mathbf{r}' \right) \otimes \left( \mathbf{r} - \mathbf{r}' \right) } {\left| \mathbf{r} - \mathbf{r}' \right|^3} \Ub_{p'} \rpp = \\
    \oint_{S} \dS \oint_{S} \dS' \left( \Ua_{p} \rp \cdot \n \right)  {\left| \mathbf{r} - \mathbf{r}' \right|} \left(  \Ub_{p'} \rpp \cdot \n'  \right) + \\
    \int_V \dV \int_V \dV' \,  \frac{\Ua_p \rp \cdot \Ub_{p'} \rpp} {\left| \mathbf{r} - \mathbf{r}' \right|} , 
    \label{eq:IntegralIdentityI}
\end{multline}
\begin{multline}
    \int_V \dV \int_V \dV' \, \Ua_{p} \rp  \frac{ \left( \mathbf{r} - \mathbf{r}' \right) \otimes \left( \mathbf{r} - \mathbf{r}' \right) } {\left| \mathbf{r} - \mathbf{r}' \right|} \Ub_{p'} \rpp = \\
    -\oint_{S} \dS \oint_{S} \dS' \left( \Ua_{p} \rp \cdot \n \right)  {\left| \mathbf{r} - \mathbf{r}' \right|^2} \left(  \Ub_{p'} \rpp \cdot \n'  \right) + \\
    -\frac{1}{2} \int_V \dV \int_V \dV' \,   {\left| \mathbf{r} - \mathbf{r}' \right|} \Ua_p \rp  \cdot \Ub_{p'} \rpp.
    \label{eq:IntegralIdentityII}
\end{multline}
Furthermore, we also use the asymptotic expansion of the transverse dyadic Green function for $(sr/c_0) \rightarrow0$
\begin{multline}
 \overleftrightarrow{G}^\perp \left(\rb;s \right) =  \overleftrightarrow{G}_0^\perp \left(\rb \right) + \\ -\frac{2}{12 \pi r} \left( \frac{s \, r}{c_0} \right)  + \frac{\left( 3 \overleftrightarrow{I} - \hat{\mathbf{r}}\otimes\hat{\mathbf{r}}\right)}{32 \pi r} \left( \frac{s \, r}{c_0} \right)^2+\mathcal{O} \left( \frac{s \, r}{c_0} \right)^3.
 \label{eq:tGreenAsym}
\end{multline}
The term $\overleftrightarrow{G}_0^\perp \left(\rb \right)$ is the static transverse dyadic Green function in the free space
\begin{equation}
\label{eq:GreenStat}
    \overleftrightarrow{G}_0^\perp \left(\rb \right)=(\overleftrightarrow{I}+\hat{\mathbf{r}}\otimes\hat{\mathbf{r}})\frac{1}{8 \pi r}.
\end{equation}

\subsection{Asymptotic expansion of ${S}_{p'p}^{a b}(\zeta)$ for a spherical particle when $\beta \ll 1$ }
\label{sec:Asymptotic}
Here we give the asymptotic expansion of the coefficients ${S}_{p'p}^{a b}(\zeta)$ for a spherical particle in the limit $\beta \ll 1$. The transverse modes of E-type couple with the longitudinal modes with the same multipolar order, whereas the transverse mode of H-type only couple with themselves.
 
The self-coupling coefficient $\mathbf{U}_p^\parallel$ for the $p-$longitudinal mode  with $p=(mnv)$ is given by
\begin{equation}
\omega_P^2\,\zeta S_{pp}^{\parallel\,\parallel} \approx \beta^2 \zeta^2\left[ R_{pp}^{\parallel\,\parallel} + (i)^{2n} \beta^{2n-1} W_{pp}^{\parallel\,\parallel} \left( \frac{\zeta}{\omega_P}\right)^{2n-1} \right]
\end{equation}
where
\begin{equation}
 \label{eq:Rparallel}
 R_{pp}^{\parallel\,\parallel} =  \frac{2(n+1)}{(3+2 n)(4 n^2-1)}
\end{equation}
and
\begin{equation}
 W_{pp}^{\parallel\,\parallel} = \frac{(n+1)}{ [(2 n+1) ! !]^{2}}.
\end{equation}

The self-coupling coefficient for the transverse modes of H-type $ \mathbf{U}_p^\perp$ with $p=(mn\,1\,lv)$ (namely, $s=1)$ is given by
\begin{multline}
\omega_P^2\,\zeta S_{pp}^{\perp \, \perp} \approx \beta^2 \zeta^2 \left[\frac{a^2}{\kappa_p} - \beta^{2}R_{pp}^{\perp\,\perp} \left( \frac{\zeta}{\omega_P} \right) ^2 \right.
\\
\left. - (i)^{2n} \beta^{2n+1}W_{pp}^{\perp\,\perp}\left( \frac{\zeta}{\omega_P} \right) ^{2n+1}\right]
\end{multline}
where
\begin{equation}
 R_{pp}^{\perp\,\perp} =\frac{2n+1}{2n-1} \left(\frac{1}{a^2\kappa_p}\right)^2
\end{equation}
and
\begin{equation}
 W_{pp}^{\perp\,\perp} =\frac{2}{\left[\left(2n-1\right)!!\right]^2} \left(\frac{1}{a^2\kappa_p}\right)^2.
\end{equation}
\\
The self-coupling coefficient of the transverse modes of E-type $ \mathbf{U}_p^\perp$ with $p=(mn\,2\,lv)$ (namely, $s=2)$ is
\begin{multline}
\omega_P^2\,\zeta S_{pp}^{\perp \perp} \approx \beta^2 \zeta^2 \left[ \frac{1}{a^2\kappa_p} - \beta^2R_{pp}^{\perp\,\perp} \left( \frac{\zeta}{\omega_P}\right)^2 \right.
\\
\left. +(i)^{2n}\beta^{2n+3}W_{pp}^{\perp\,\perp}\left( \frac{\zeta}{\omega_P} \right) ^{2n+3} \right]
\end{multline}
where
\begin{equation}
 R_{pp}^{\perp\,\perp} =\frac{2n+3}{2n+1} \left(\frac{1}{a^2\kappa_p}\right)^2
\end{equation}
and
\begin{equation}
 W_{pp}^{\perp\,\perp} =\frac{2}{\left[ \left( 2n + 1 \right)!! \right]^2} \left(\frac{1}{a^2\kappa_p}\right)^2.
\end{equation}

At last, the mutual coupling coefficient between the transverse mode of E-type $ \mathbf{U}_p^\perp$ with $p=(mn\,2\,lv)$ (namely, $s=2$) and the longitudinal mode $ \mathbf{U}_{p'}^\parallel$ with $p'=mnv$ is given by
\begin{equation}
\omega_P^2\,\zeta S_{pp'}^{\perp \parallel}  \approx \beta^2 \zeta^2\left[R_{pp'}^{\perp\,\parallel}  - (i)^{2n} \beta^{2n+1} W_{pp'}^{\perp\,\parallel} \left(\frac{\zeta}{\omega_P} \right) ^{2n+1}\right]
\end{equation}
where
\begin{equation}
 R_{pp'}^{\perp\,\parallel} =\frac{1}{z_{n,l}^2} \frac{\sqrt{2 \left( n + 1 \right)}}{ \left( 2 n + 1 \right)}
\end{equation}
and
\begin{equation}
 W_{pp'}^{\perp\,\parallel} =\frac{1}{z_{n,l}^2} \frac{2 \left( n + 1 \right)}{\left[ \left( 2n + 1 \right)!! \right]^2 \sqrt{2 \left( n + 1 \right)}}.
\end{equation}
These asymptotic expansions allow us to evaluate the frequency shift and the decay rate of the natural modes of the polarization field in the small size limit $a \ll c_0/\omega_P$ given in Section VII.A.

\end{document}